\documentclass[preprint,12pt]{elsarticle}

\usepackage{amssymb}
\usepackage{amsmath}
\usepackage{bm}
\usepackage{xcolor}
\usepackage[caption=false,listofformat=subsimple,labelformat=simple]{subfig}
\usepackage{adjustbox}

\usepackage{hyperref}
\hypersetup{
	colorlinks=true,
	linkcolor=blue,
	urlcolor=cyan,
	}

\biboptions{sort&compress}
\journal{Chinese Journal of Physics}

\begin{document}

\begin{frontmatter}



\title{Quaking in Soft Granular Particles with Speed-dependent Friction: Role of Critical Volume Fraction and Inertia} 


\author{Wei-Chang Lo} 

\author{Jih-Chiang Tsai} 

\affiliation{organization={Institute of Physics, Academia Sinica},
            city={Taipei},
            postcode={115201}, 
            country={Taiwan}}

\begin{abstract}
Our previous numerical simulation [C.-E. Tsai et al., Physical Review Research \textbf{6}, 023065 (2024)] has shown that, for soft granular particles under quasistatic shearing, incorporating a speed-dependent friction is essential to reproduce the rate-dependent stick-slip fluctuations that have been found in the laboratory experiment [J.-C. Tsai et al., Physical Review Letters \textbf{126}, 128001 (2021)]. As a continuation, here we employ the simulation in a wide range of driving speeds to examine the role of grain inertia in the quaking dynamics. With our Stribeck-Hertz model, we find that having the volume fraction exceeding a critical value $\phi_{\text{c}}$ is a necessary condition for the quaking to occur, and that the value of $\phi_{\text{c}}$ is determined by material parameters only, independent of the driving rate. The effect of grain inertia generally suppresses the occurrence of quaking, and we conclude by presenting the state diagrams which exhibit a progressive narrowing of the quaking regime as the driving speed increases and the disappearance of quaking at an extremely high shear rate.
\end{abstract}



\begin{keyword}
granular materials \sep slip avalanches \sep inertial rheology


\end{keyword}

\end{frontmatter}


\section{Introduction}
Various seismological views on the mechanisms of earthquake and fault dynamics have progressively developed in the past century, including the recognition of fractality at the fault zone and granular aspects of earthquake and fault dynamics \cite{BenZion2008}. Therefore, the intermittent dynamics (e.g. slip avalanches) in granular materials has been extensively investigated to inspect whether it reproduces the scale-invariant power laws in seismology \cite{Bretz2006,Dahmen2011,deArcangelis2016,Ma2020,Sultan2022,Petri2024}. A great deal of such studies were performed using dry granular particles. However, fluid content at the fault zone is one of the main factors that leads to deviation of magnitude-frequency distributions from the seminal Gutenberg-Richter law \cite{Ben-Zion2006,BenZion2008}. Meanwhile, there is evidence that faults are lubricated during earthquakes \cite{DiToro2011}. Velocity weakening in fault gouge and in induced earthquakes via fluid injection have also been reported \cite{Mizoguchi2009,Wang2020,Rubino2022}. These works have raised our interest into the effect of lubrication upon the intermittency in a sheared granular material.

We previously conducted a laboratory experiment of granular shear flow with an interstitial fluid that exhibited strong intermittency, characterized by bursts of grain-level displacement and sudden release of stresses only in the mid-range of the driving rates \cite{Tsai2021}. To understand this intriguing rate dependence, we have also investigated the tribology between surfaces of these granular particles immersed in the same fluid \cite{Tsai2024a}: The time averaged friction force shows a nonmonotonic dependence on the sliding speed between the particles that is consistent with the paradigmatic Stribeck diagram \cite{Jacobson2003,Williams2005}. More specifically, at low speeds, we have identified an onset value $V_{\text{c}}$ beyond which the friction between particles begins to reduce, presumably due to the lubrication of the interstitial fluid. Based on these empirical facts, we have also set up a numerical simulation \cite{Tsai2024b} to demonstrate that, for a granular packing under continuous shearing, the speed-dependent friction is not only a sufficient but also a necessary condition for the sudden changes of several physical quantities, including the coordination number, the elastic energy, and the boundary stress. Collectively, we call these sudden changes \emph{quaking} for its resemblance to earthquakes. In the simulation, we have used a dimensionless parameter called the \emph{slipperiness}, $\dot{\gamma}d/V_{\text{c}}$, as one of the main control parameters (in addition to the volume fraction $\phi$). Here, the physical shear rate is defined as $\dot{\gamma}\equiv U/Z_0$, where $U$ is the driving speed, $Z_0$ is the thickness of the granular packing, $d$ stands for the mean diameter of particles and $V_{\text{c}}$ for the aforementioned onset speed of the velocity weakening.

In our previous work \cite{Tsai2024b}, in order to focus on quasistatic flows, $U$ was limited to a low value (0.1 cm/s). Several decades of changes in the value of $\dot{\gamma}d/V_{\text{c}}$ were achieved by varying the material parameter $V_{\text{c}}$ hypothetically. However, a more natural, real-world like approach should have been maintaining the material parameter $V_{\text{c}}$ and varying the driving speed $U$ instead. One important motivation of the present work is to go beyond the quasistatic limit and understand whether increasing the driving speed $U$ and, in turn, bringing in the effect of inertia would change the quaking dynamics.

Going beyond the quasistatic limit also means the substantial reduction of packing densities below the critical point of transition. In the present work, we locate the critical points of transition in the steady-shear dynamics, and use these critical points as the baseline to compare models with or without speed-dependent friction.


\section{Methodology}
\subsection{Numerical setup}
As in our previous work, our simulations are carried out in LAMMPS \cite{Thompson2022}---please see Ref.~\cite{Tsai2024b} for details of our simple-shear geometry, the binary mixture of soft spheres in two diameters (0.9$d_{\text{w}}$ and 1.1$d_{\text{w}}$, in which $d_{\text{w}}$ stands for the diameter of particles constituting the driving walls) with fixed number for both, the driving walls moving along $\pm X$ directions at a constant velocity $\pm U/2$ and a fixed separation $Z_0$ for each target volume fraction, the periodic boundary conditions in the direction of $X$ and $Y$, and the material parameters (elastic moduli and mass) mimicking our laboratory experiment \cite{Tsai2021}. Here we merely describe the essentials. In most cases, we make a side-by-side comparison between the results from two models as follows:
\begin{enumerate}
	\item The Coulomb-Hertz (CH) model, also known as the ``history-dependent Hertz model'' in LAMMPS' documentation \cite{lammps}. In this model, particles $i$ and $j$ interact viscoelastically via the \emph{overlap distance} $\delta_{ij}=R_i+R_j-r_{ij}$ as the strain of deformation, where $R_i$ and $R_j$ are the particle radii, and $r_{ij}$ is the separation between the centers of these two particles; note that $\delta_{ij}\equiv 0$ if $r_{ij}>R_i+R_j$. In the normal direction, the force $\bm{f}_{\text{N}}^{ij}$ follows the law of a Hertzian contact with a viscous damping:
\begin{equation}\label{eq:normal_force}
	\bm{f}_{\text{N}}^{ij}=\sqrt{\delta_{ij}R_{\text{eff}}}(K_{\text{N}}\delta_{ij}\bm{\hat{n}}_{ij}-m_{\text{eff}}\gamma_{\text{N}}\bm{v}_{\text{N}}^{ij}),
\end{equation}	
where $R_{\text{eff}}$, $m_{\text{eff}}$, $\bm{v}_{\text{N}}^{ij}$ and $\bm{\hat{n}}_{ij}$ are the effective radius, the effective mass, the normal component of the relative velocity and the unit vector along the line connecting the centers of the two particles, respectively, and $K_{\text{N}}$ is the elastic constant and $\gamma_{\text{N}}$ is the viscoelastic damping constant for normal contact, respectively \cite{lammps}. In the tangential direction, the force without damping,
\begin{equation}\label{eq:tangential_force}
	\bm{f}_{\text{T}}^{ij}=-\sqrt{\delta_{ij}R_{\text{eff}}}K_{\text{T}}\Delta\bm{S}_{\text{T}}^{ij},
\end{equation}
grows linearly with the lateral movement between particle surfaces $\Delta\bm{S}_{\text{T}}^{ij}$ since the moment of contact, up to the saturation magnitude $f_{\text{T}}^{ij}=\mu_0f_{\text{N}}^{ij}$, until the two spheres fall apart, i.e., $\delta_{ij}=0$. The Coulomb coefficient $\mu_0$ and the elastic constant for tangential contact $K_{\text{T}}$ are both material constants that are independent of particle movement. In this work, we set $K_{\text{N}}=K_{\text{T}}=1.5$ MPa, and $\gamma_{\text{N}}=1.5\times 10^5$ (cm$\cdot$s)${}^{-1}$ by default unless otherwise specified. The choice of these parameters is based on the previous experimental characterization of the PDMS spheres \cite{Tsai2021}.
	\item The Stribeck-Hertz (SH) model. The interparticle force laws are the same as the CH model except a slight modification: To replicate the velocity weakening reported in Ref.~\cite{Tsai2024a}, we replace the constant $\mu_0$ with a speed-dependent friction coefficient
	\begin{equation}\label{eq:stribeck-curve}
	\mu(v_{\text{T}}) =
  		\begin{cases}
    	\mu(0)       & \quad \text{if } v_{\text{T}} \le V_{\text{c}}\\
     	\mu(0)\times(v_{\text{T}}/V_{\text{c}})^{-1} & \quad \text{if }v_{\text{T}} > V_{\text{c}},
  		\end{cases}
	\end{equation}
	where $v_{\text{T}}$ stands for the sliding speed between two particles when they are overlapping, that is, the magnitude of time derivatives of $\Delta\bm{S}_{\text{T}}^{ij}$, and $V_{\text{c}}$ is a material-specific characteristic speed that corresponds to the onset of the velocity weakening. As a continuation of our previous work, the other material-specific constant $\mu(0)$ is set to unity.
\end{enumerate}

\subsection{Identifying the critical volume fraction}
Since the magnitude of the tangential force in the SH model is capped at $f_{\text{T}}^{ij}=\mu(v_{\text{T}})f_{\text{N}}^{ij}$, and $\mu(v_{\text{T}})$ is floating according to the magnitudes of $v_{\text{T}}$ and $V_{\text{c}}$, the maximum of $f_{\text{T}}^{ij}/f_{\text{N}}^{ij}$ is now floating as well. In other words, the distribution of $f_{\text{T}}^{ij}/f_{\text{N}}^{ij}$ mixes (i) those are yet to reach the threshold $\mu(0)$ and (ii) those are weakened down to $\mu(0)\times(v_{\text{T}}/V_{\text{c}})^{-1}$. Measuring $f_{\text{T}}^{ij}/f_{\text{N}}^{ij}$ or the (global) stress ratio would not necessarily reflect the actual interparticle friction coefficient, which is a crucial quantity for examining the dynamics of granular flows. Hence, we aim to circumvent this issue and adopt an experimentally accessible approach to investigate the dynamics governed by the SH model.

Generally speaking, the dynamics of granular particles under an imposed shear is controlled by the volume fraction $\phi$ and the shear rate $\dot{\gamma}$. These granular particles start to form a network of enduring contacts as the volume fraction $\phi$ reaches a critical value $\phi_{\text{c}}$ \cite{Ji2008,Vescovi2016}. For infinitely rigid spheres, $\phi_{\text{c}}$ corresponds to the jamming point \cite{Aharonov1999,Song2008}. For soft particles, on the other hand, a volume fraction beyond $\phi_{\text{c}}$ is accessible. With the increased volume fraction, the deformation at the contacts become progressively persistent and dominate the rheological responses. The critical point $\phi_{\text{c}}$ has been used as the reference of transition in comparing the rheology of debris flow samples \cite{Kostynick2022}. Specifically, the critical volume fraction of soft granular particles represents a regime transition between collisional flows ($\phi<\phi_{\text{c}}$) and elastoplastic deformation ($\phi>\phi_{\text{c}}$); the former exhibits a dominant effect of particle inertia, and the latter is considered quasistatic \cite{Chialvo2012}. It is expected that the granular dynamics governed by the CH or the SH model would exhibit substantial differences only when $\phi>\phi_{\text{c}}$, such that the role of interparticle friction becomes significant because of the enduring overlaps.

\begin{figure*}
	\subfloat[$\mu_0=0.001$\label{fig:xi_Z_mu0low}]{
		\includegraphics[width=0.33\textwidth]{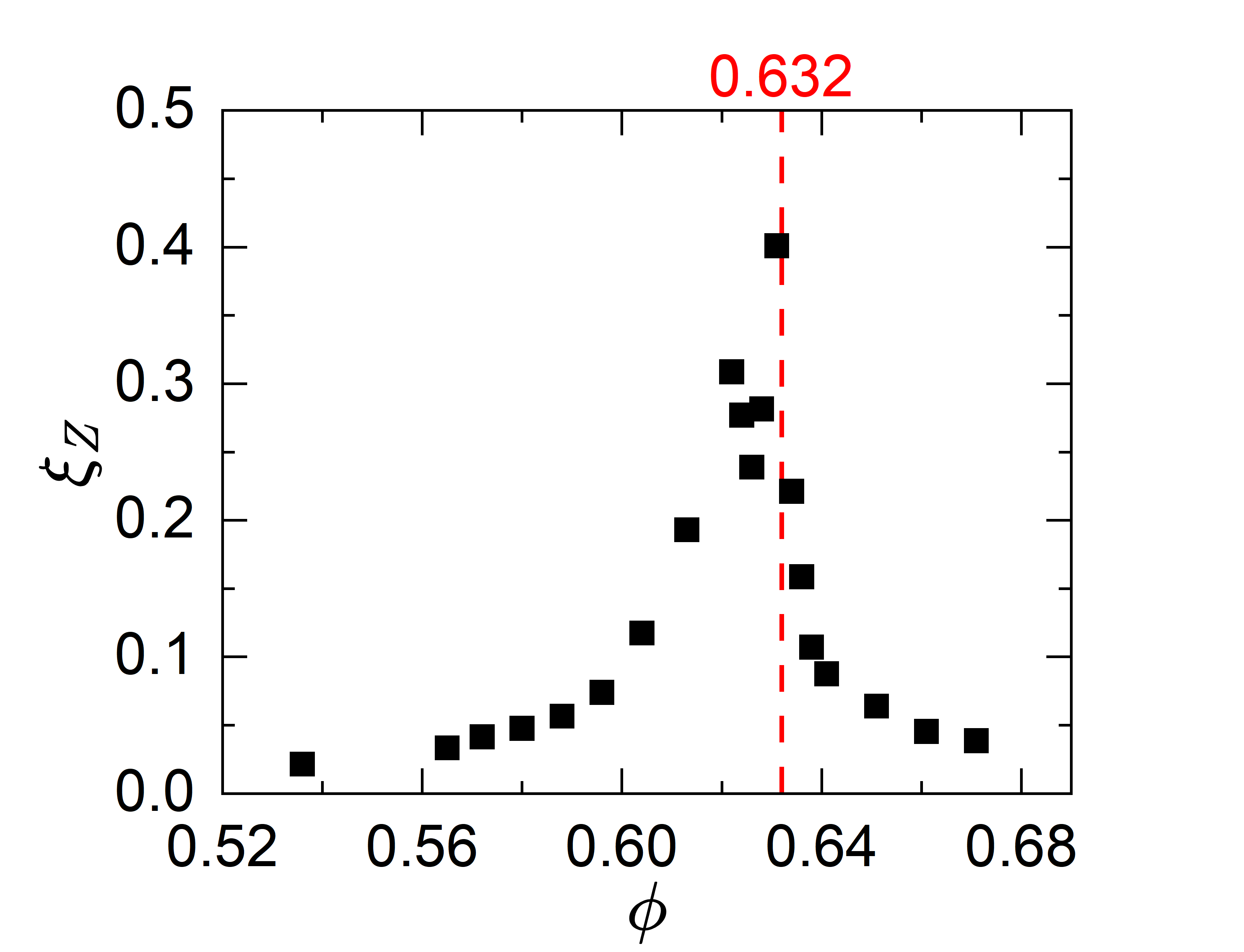}
	}\hfill
	\subfloat[$V_{\text{c}}=6\times 10^{-6}$ cm/s\label{fig:xi_Z_Vclow_slow}]{
		\includegraphics[width=0.33\textwidth]{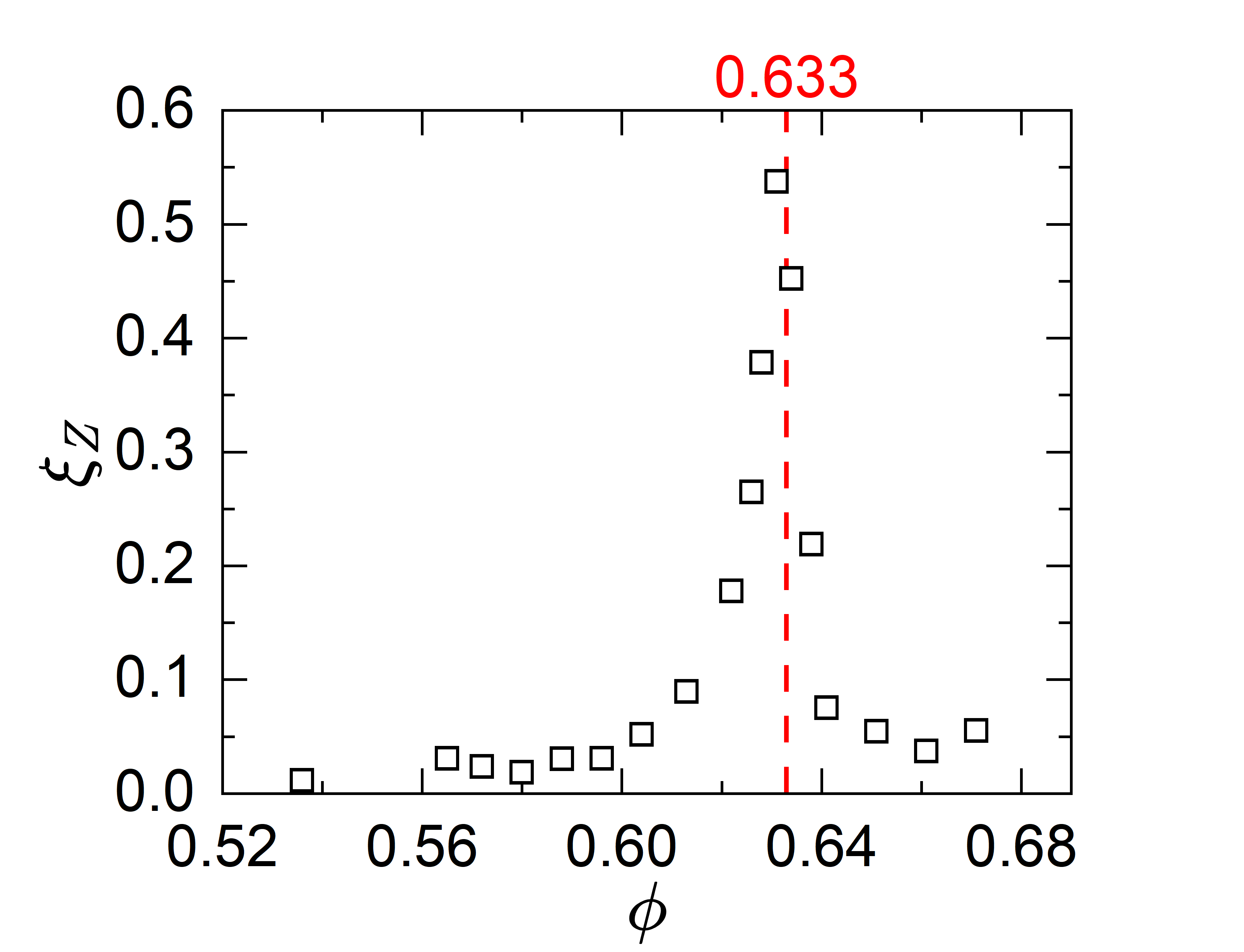}
	}
	\subfloat[$V_{\text{c}}=6\times 10^{-4}$ cm/s\label{fig:xi_Z_Vclow_fast}]{
		\includegraphics[width=0.33\textwidth]{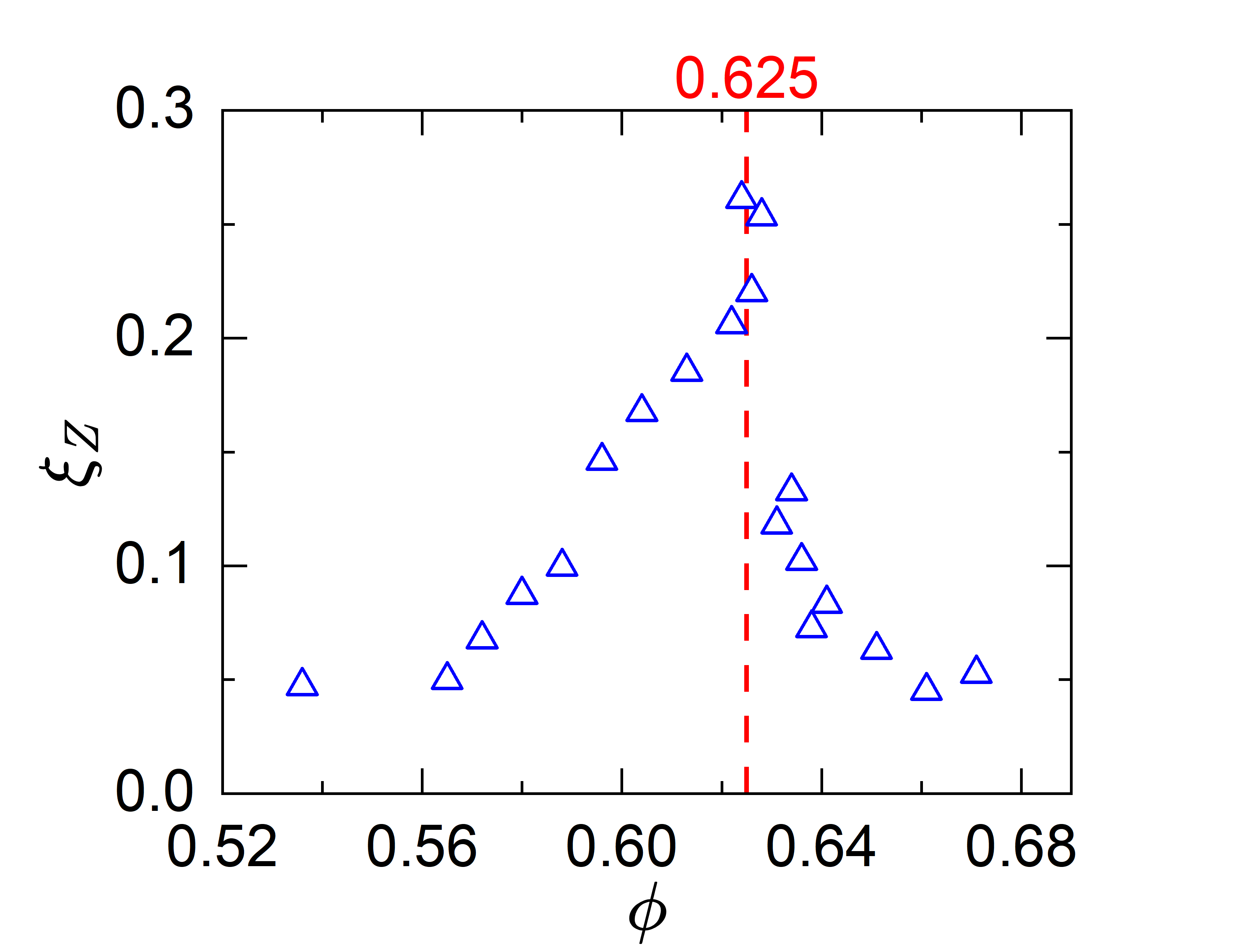}
	}
	
	\subfloat[$\mu_0=0.2$\label{fig:xi_Z_mu0mid}]{
		\includegraphics[width=0.33\textwidth]{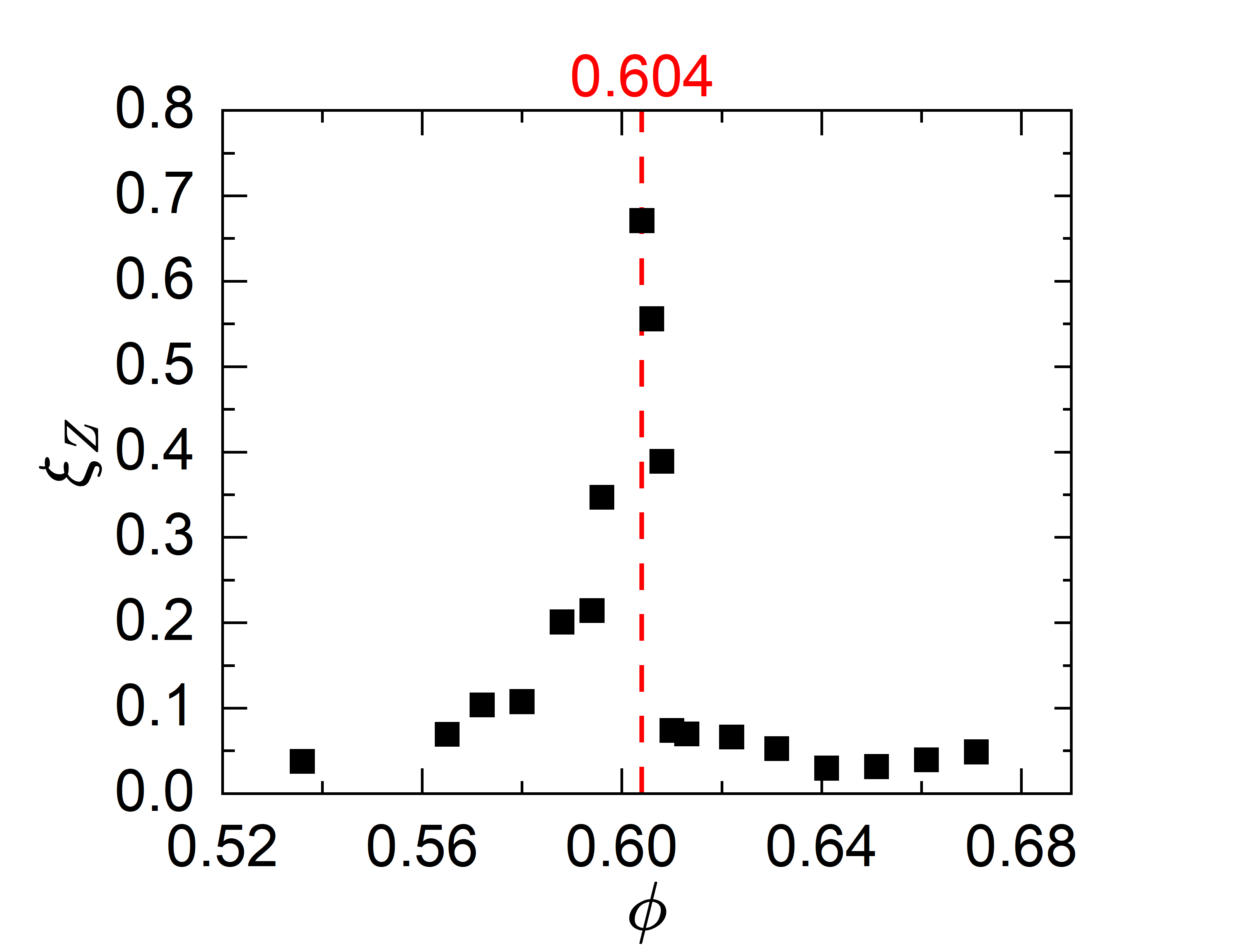}
	}\hfill
	\subfloat[$V_{\text{c}}=0.06$ cm/s\label{fig:xi_Z_Vcmid_slow}]{
		\includegraphics[width=0.33\textwidth]{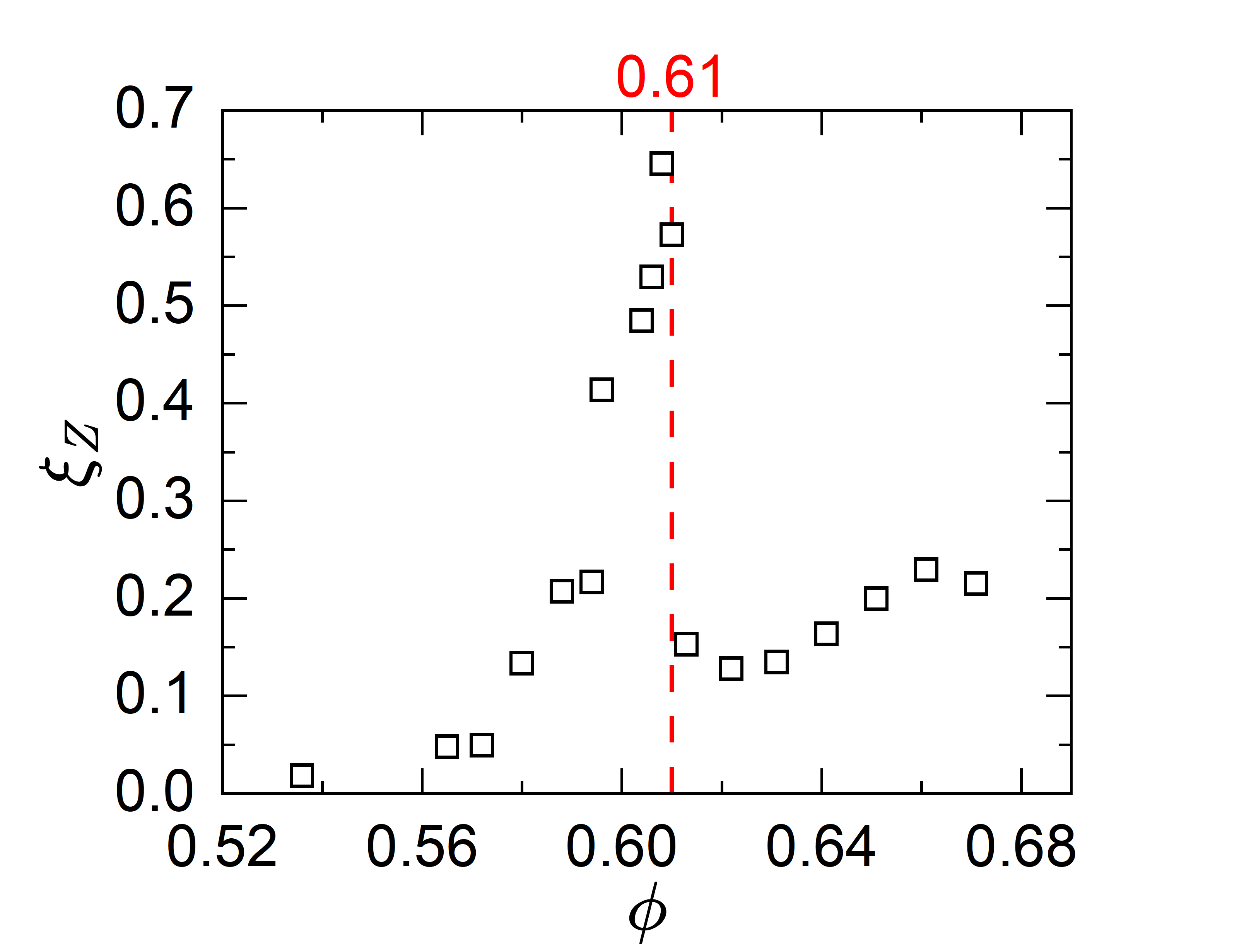}
	}
	\subfloat[$V_{\text{c}}=0.06$ cm/s\label{fig:xi_Z_Vcmid_fast}]{
		\includegraphics[width=0.33\textwidth]{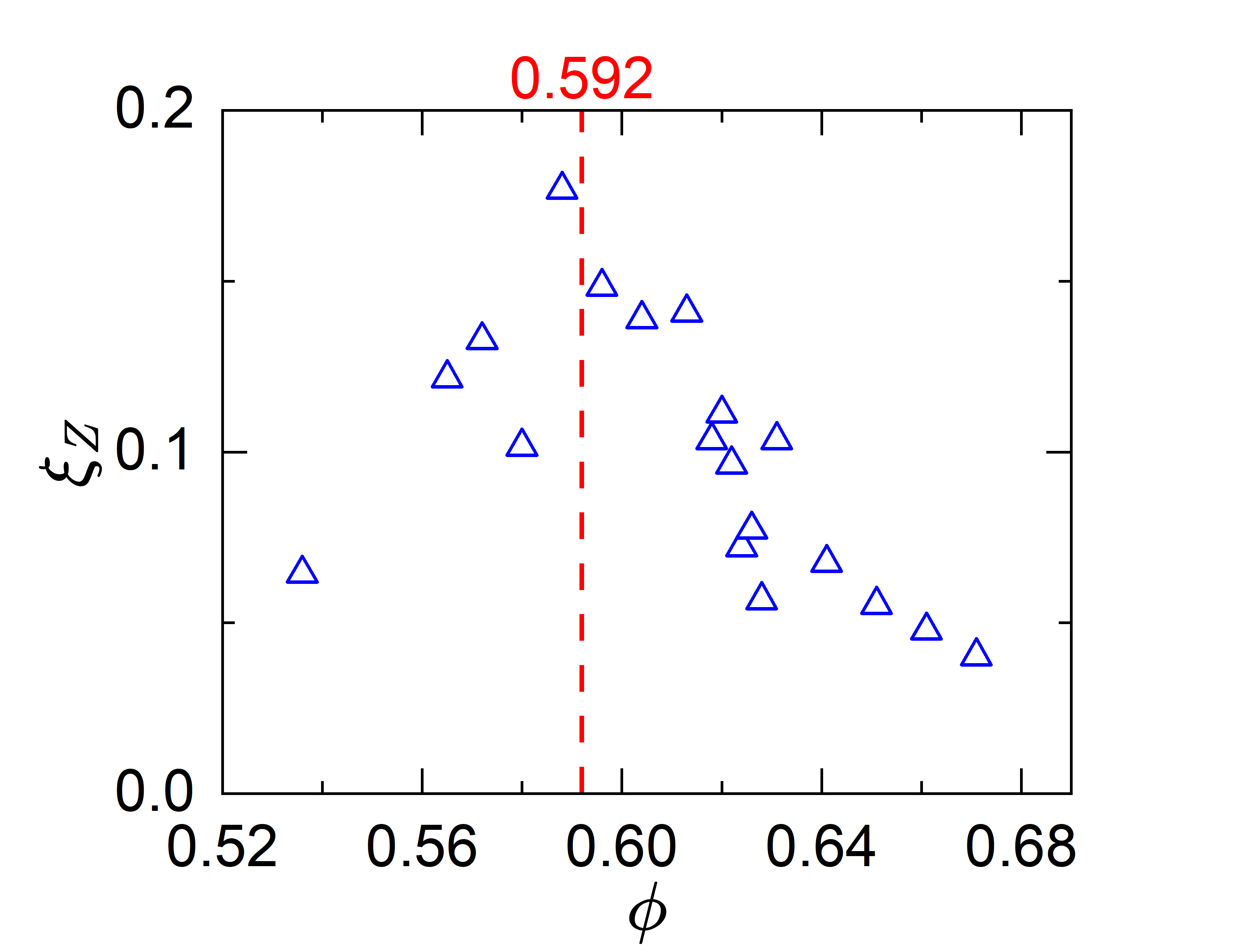}
	}
	
	\subfloat[$\mu_0=1.0$\label{fig:xi_Z_mu0high}]{
		\includegraphics[width=0.33\textwidth]{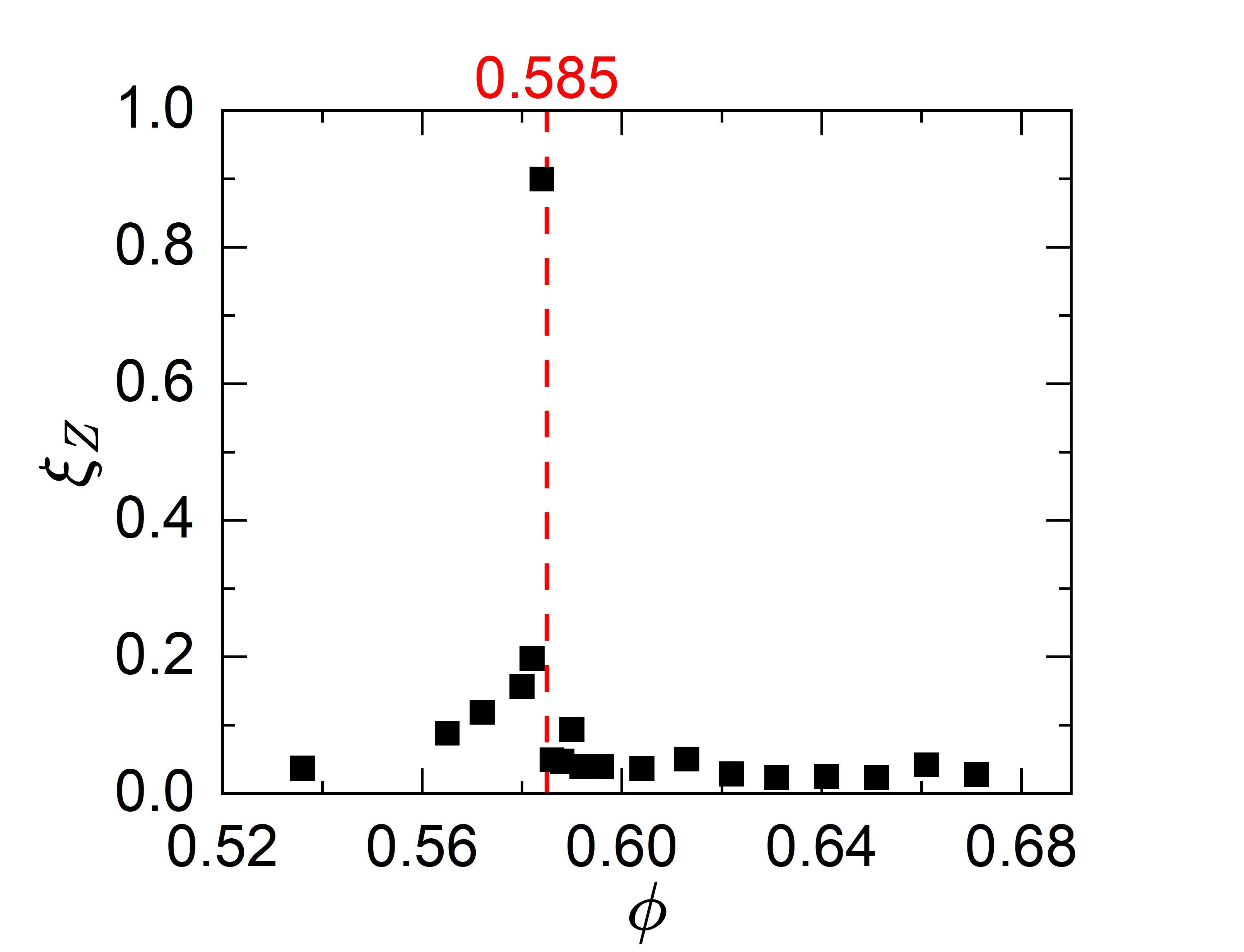}
	}\hfill
	\subfloat[$V_{\text{c}}=600$ cm/s\label{fig:xi_Z_Vchigh_slow}]{
		\includegraphics[width=0.33\textwidth]{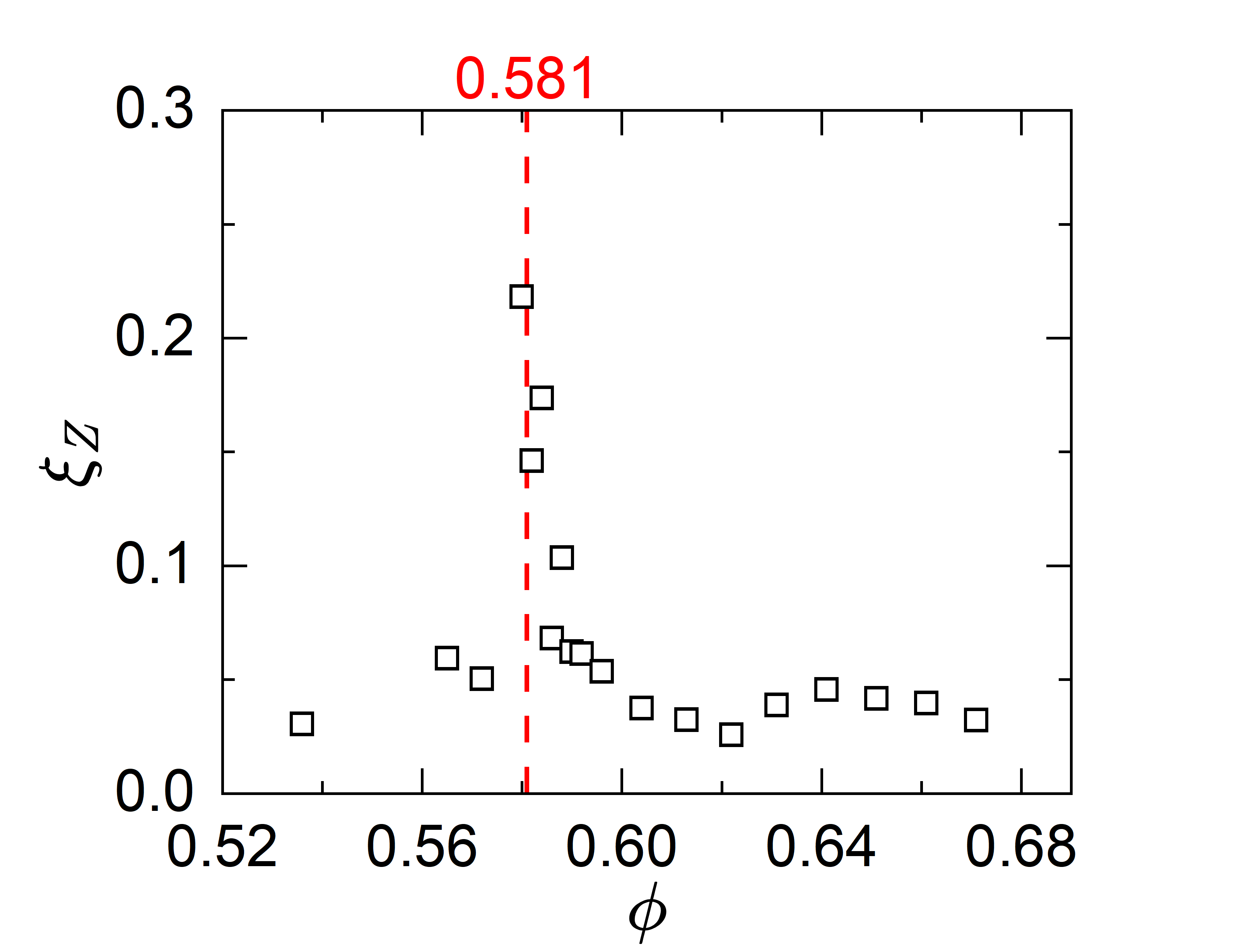}
	}
	\subfloat[$V_{\text{c}}=6\times 10^4$ cm/s\label{fig:xi_Z_Vchigh_fast}]{
		\includegraphics[width=0.33\textwidth]{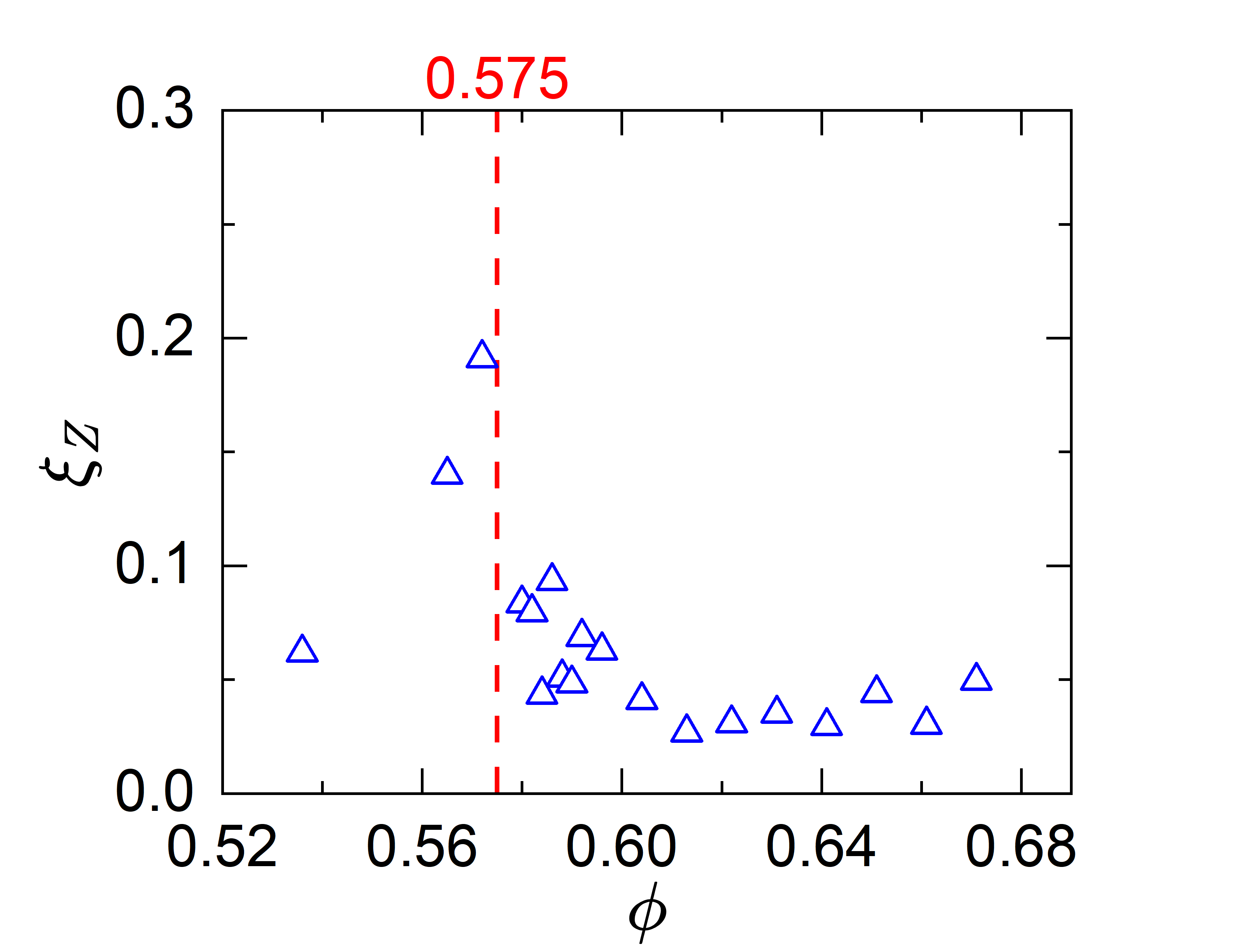}
	}
	\caption{Nine examples of the fluctuation of coordination numbers $\xi_Z$ that varies with the volume fraction $\phi$, and with various Coulomb coefficients $\mu_0$ (the CH model) or weakening onset speeds $V_{\text{c}}$ (the SH model). The left column shows examples from the CH model. The middle and the right columns show examples from the SH model with $U=0.01$ cm/s and $U=1.0$ cm/s, respectively. The vertical dashed line indicates the value of the critical volume fraction for each panel.}\label{fig:find_phic}
\end{figure*}

A number of ways to locate $\phi_{\text{c}}$ of soft particles have been discussed in the literature; see Ref.~\cite{Berzi2024} and the references therein. In general, these methods usually involve identifying the diverged fluctuations that are associated with building up the contact network by increasing $\phi$. Here we choose to look into the fluctuation of the coordination number $Z(t)$, which is the contact numbers per particle averaged over all particles (excluding the wall particles) at time $t$. For each value of $\phi$, the fluctuation of the coordination number is represented by the standard deviation $\xi_Z(\phi)=\sqrt{\sum_{\ell=1}^n[Z(\phi, t_{\ell})-\bar{Z}]^2/(n-1)}$, where $\bar{Z}$ is the mean of $Z(\phi, t_{\ell})$, $t_{\ell}$ is the sampled time point and $n$ is the number of samples. In other words, $\xi_Z$ estimates the spread of the contact number of individual particles relative to the mean of the population packed and sheared at a given volume fraction $\phi$. Figure~\ref{fig:find_phic} shows a drastic increase of $\xi_Z$ around a narrow range of $\phi$ in most of the examples. We thus calculated the finite difference of $\xi_Z$ and interpolated where the slope changed sign as the critical volume fraction $\phi_{\text{c}}$, indicated by the vertical dashed lines in Fig.~\ref{fig:find_phic}.

\section{Main Results}
\subsection{Critical volume fraction depending on material parameters}
\label{sec:phi_c}

\begin{figure*}
	\subfloat[CH model\label{fig:phic-Vc-a}]{%
  		\includegraphics[width=0.5\textwidth]{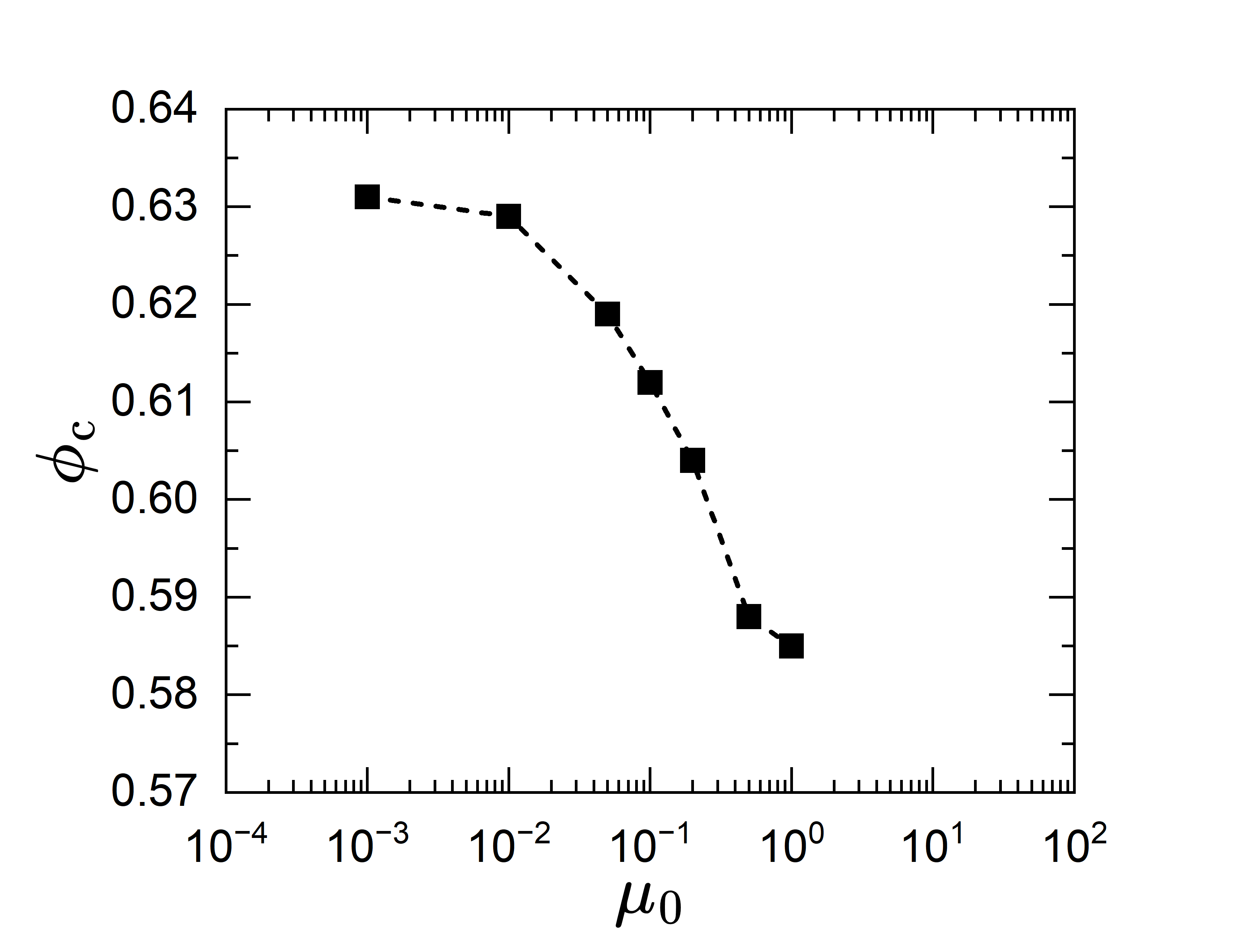}%
	}\hfill
	\subfloat[SH model\label{fig:phic-Vc-b}]{%
  		\includegraphics[width=0.5\textwidth]{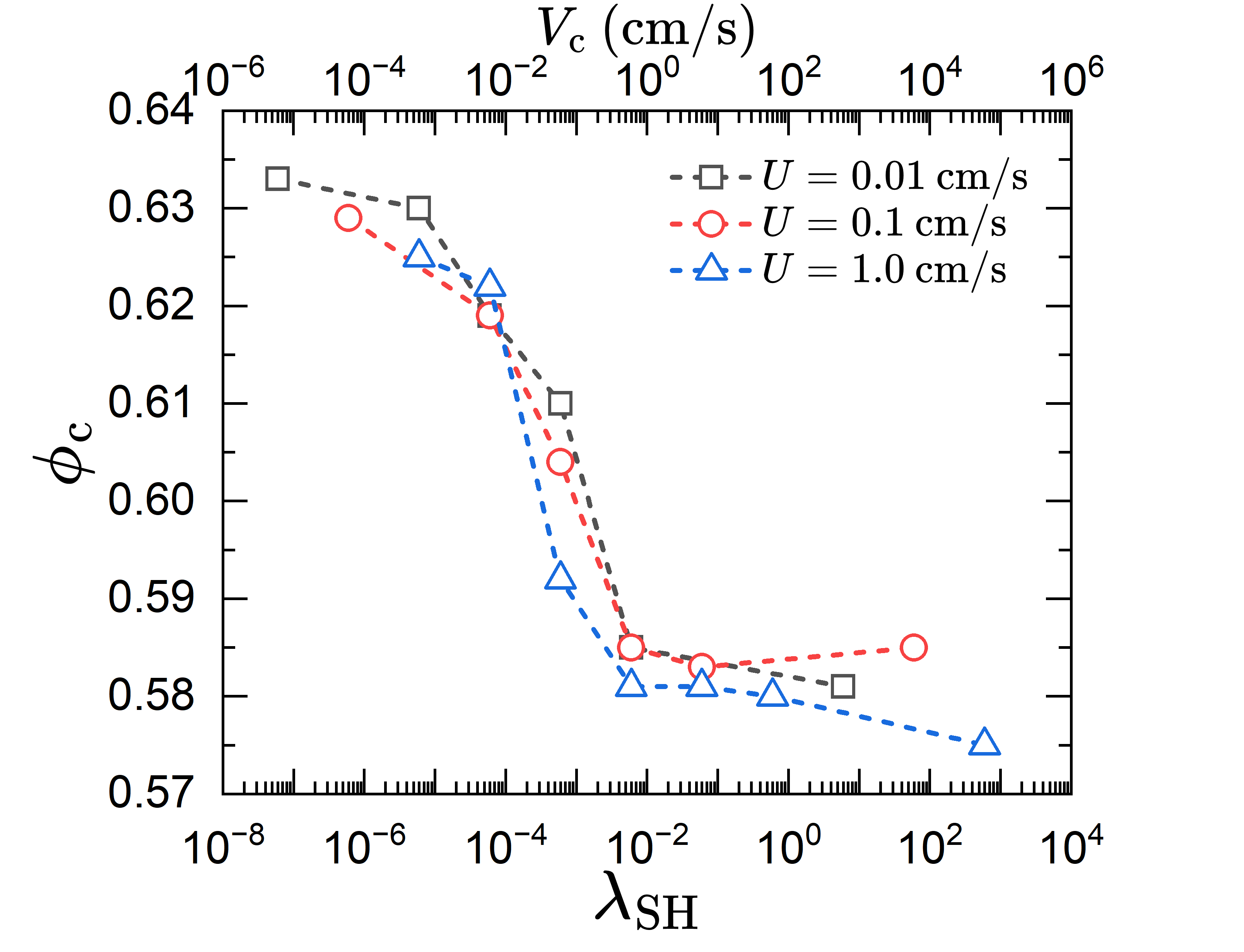}%
	}
	\caption{The critical volume fraction $\phi_{\text{c}}$ versus material-specific parameters. (a) $\phi_{\text{c}}$ versus the Coulomb coefficient $\mu_0$ in the CH model, with $U=0.1$ cm/s. (b) $\phi_{\text{c}}$ versus a new dimensionless parameter, $\lambda_{\text{SH}}=m\gamma_{\text{N}}V_{\text{c}}/K_{\text{N}}d$, with three different driving speeds. In these simulations, $K_{\text{N}}=1.5$ MPa and $\gamma_{\text{N}}=1.5\times 10^5$ (cm$\cdot$s)${}^{-1}$ were fixed, and we varied $\lambda_{\text{SH}}$ by changing $V_{\text{c}}$.}\label{fig:phic-Vc}
\end{figure*}

Figure~\ref{fig:phic-Vc-a} reproduces the well known fact that the critical volume fraction $\phi_{\text{c}}$ in the conventional (CH) model drops monotonically with the increasing value of $\mu_0$. It has been reported that the values of $\phi_{\text{c}}$ depend on $\mu_0$ but not on the shear rate $\dot{\gamma}$ \cite{Chialvo2012}. These values are also confined by two previously reported limits: $\phi_{\text{c}}\approx 0.64$ while $\mu_0\to 0$, and $\phi_{\text{c}}\approx 0.55$ while $\mu_0\to\infty$ \cite{Berzi2024}.

Interestingly, in our SH model, Fig.~\ref{fig:phic-Vc-b} shows that the value of $\phi_{\text{c}}$ is also monotonically dependent on the material-specific parameter $V_{\text{c}}$ but not sensitive to the change of driving speed $U$ by two orders of magnitude. The trend of $\phi_{\text{c}}$ in Fig.~\ref{fig:phic-Vc-b} are consistent with two \emph{trivial} limits. For $V_{\text{c}}\rightarrow\infty$, the particles are never weakened thus $\mu(v_{\text{T}})=\mu(0)=1$; in this limit $\phi_{\text{c}}$ should be the same as $\mu_0=1$ in the CH model, i.e., the rightmost point in Fig.~\ref{fig:phic-Vc-a} where $\phi_{\text{c}}=0.585$. On the other hand, when $V_{\text{c}}\rightarrow 0$, even the slightest sliding shall trigger the weakening, and the particles behave like they are frictionless; in this limit, $\phi_{\text{c}}$ should be asymptotically approaching 0.64, but the minimal $V_{\text{c}}$ and the driving speeds we have tested lead to $\phi_{\text{c}}$ close to that corresponding to $\mu_0=0.001$ in the CH model, i.e., the leftmost point in Fig.~\ref{fig:phic-Vc-a} where $\phi_{\text{c}}=0.632$.

\begin{figure}
	\includegraphics[width=\columnwidth]{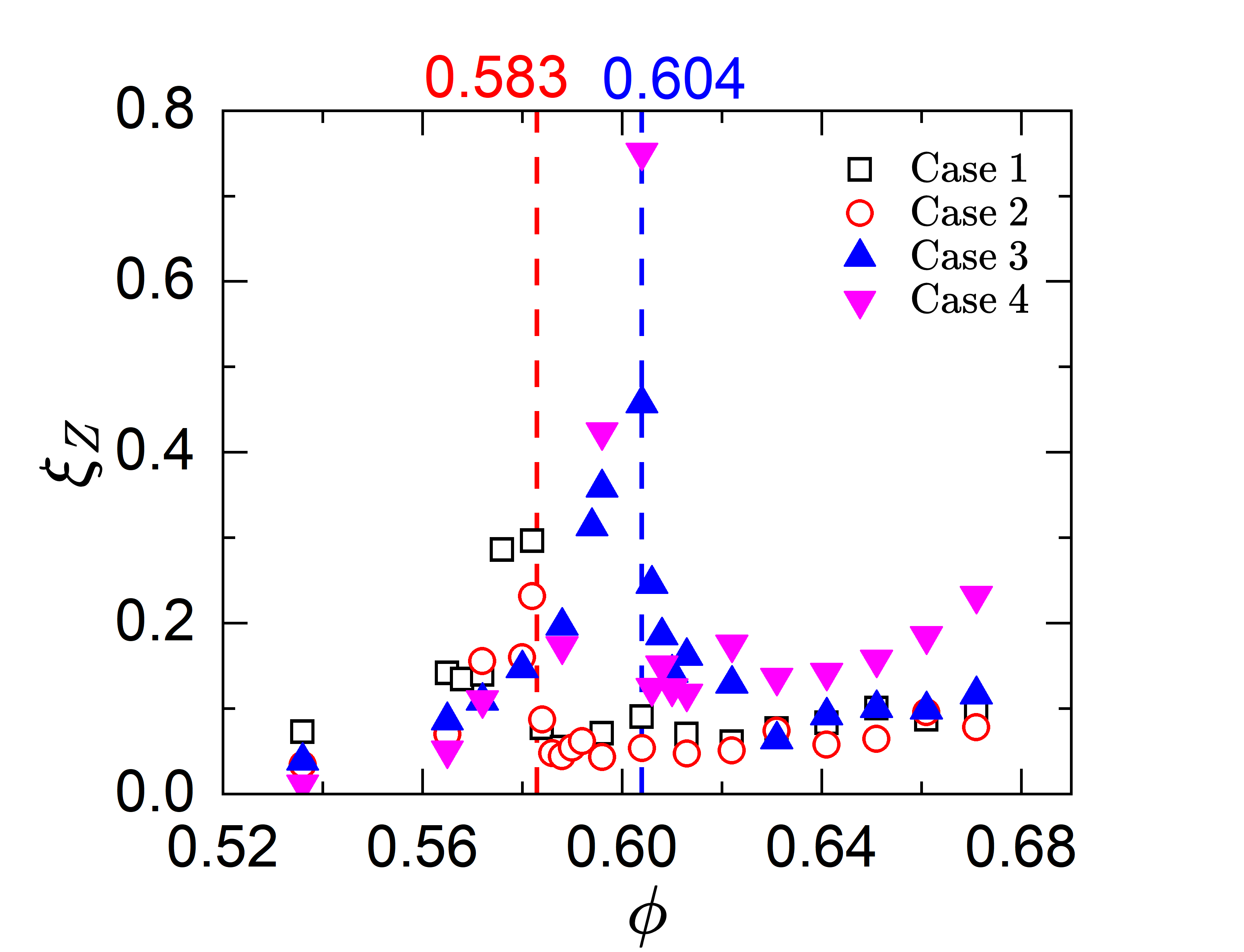}
	\caption{Simulations governed by the SH model with $\lambda_{\text{SH}}=6\times 10^{-2}$ (Case 1 and 2) and $\lambda_{\text{SH}}=6\times 10^{-4}$ (Case 3 and 4) are characterized by $\phi_{\text{c}}=0.583$ and $\phi_{\text{c}}=0.604$, respectively. Note that these simulations had various combination of parameters as shown in Table~\ref{table:multi-lambda}.}\label{fig:Z-phi_lambda}
	
\end{figure}

\begin{table}
\begin{center}
\begin{tabular}{|c|| c| c| c|| c|} 
 \hline
 Case & $V_{\text{c}}$ (cm/s) & $K_{\text{N}}$ (MPa) & $\gamma_{\text{N}}$ (cm$\cdot$s)${}^{-1}$ & $\lambda_{\text{SH}}$\\ [0.5ex] 
 \hline\hline
 1 & 0.6 & 0.15 & $1.5\times 10^5$ & $6\times 10^{-2}$\\ 
 \hline
 2 & 6.0 & 1.5 & $1.5\times 10^5$ & $6\times 10^{-2}$\\
 \hline
 3 & 0.06 & 1.5 & $1.5\times 10^5$ & $6\times 10^{-4}$\\
 \hline
 4 & 0.006 & 15.0 & $1.5\times 10^7$ & $6\times 10^{-4}$\\
 \hline
\end{tabular}
\caption{Parameters used in the simulations that are plotted in Fig.~\ref{fig:Z-phi_lambda}.}
\label{table:multi-lambda}
\end{center}
\end{table}

The dependence on the material parameters shown in Fig.~\ref{fig:phic-Vc-b} can be nondimensionalized as we define $\lambda_{\text{SH}}\equiv m\gamma_{\text{N}}V_{\text{c}}/K_{\text{N}}d$, where $m$ is the mean mass of the particles. We may interpret $\lambda_{\text{SH}}$ as a ratio of the retardation time $\tau_{\text{R}}=m\gamma_{\text{N}}/K_{\text{N}}$ to the critical weakening time $\tau_{\text{c}}=d/V_{\text{c}}$. The sliding velocity between two colliding particles is coupled with how long they stay overlapped while increasing the tangential displacement $\Delta\bm{S}_{\text{T}}^{ij}$ in Eq.~(\ref{eq:tangential_force}). Hence, $\lambda_{\text{SH}}$ can be seen as a measure of whether the duration of interparticle sliding while transiently overlapping is short enough to trigger the velocity weakening; the physical meaning of $\lambda_{\text{SH}}$ will be further elaborated in Sec.~\ref{sec:discussion}.

To show the validity of $\phi_{\text{c}}$ being a function of $\lambda_{\text{SH}}$, we plot the fluctuation of coordination numbers $\xi_Z$ of four examples with varied parameters $V_{\text{c}}$, $K_{\text{N}}$ and $\gamma_{\text{N}}$ that correspond to two different values of $\lambda_{\text{SH}}$ (Table \ref{table:multi-lambda}) in Figure~\ref{fig:Z-phi_lambda}. It is evident that various combination of these parameters which correspond to the same value of $\lambda_{\text{SH}}$ will result in the same value of $\phi_{\text{c}}$.

Overall, $\phi_{\text{c}}$ as a function of $\lambda_{\text{SH}}$ in Fig.~\ref{fig:phic-Vc-b} qualitatively resemble those as a function of $\mu_0$ in Fig.~\ref{fig:phic-Vc-a}. The value of $\phi_{\text{c}}$ offers a convenient basis for comparing simulations governed by the CH and the SH models. Given that the values of $\phi_{\text{c}}$ for the CH model and for the SH model are both material-specific and insensitive to the driving speed, we speculate that there would be a mapping between the dynamics governed by the CH and the SH model via the distance to a similar value of $\phi_{\text{c}}$. We then will have a convenient baseline to group the data from both models to make side-by-side comparisons. In the following two subsections, detailed comparisons are made based on the choice of material parameters with which both models share a similar value of $\phi_{\text{c}}$.

\subsection{Verifying universal behaviors disclosed by the distance to $\phi_{\text{c}}$}
The steady-state dynamics governed by both models, at various shear rates tested in our simulations, quantitatively falls into \emph{two} distinct regimes based on the difference between $\phi$ and $\phi_{\text{c}}$: collisional flows ($\phi<\phi_{\text{c}}$) and quasistatic flows ($\phi>\phi_{\text{c}}$). Each regime features a certain scaling law between variables, for instance, the normal stress and the shear rate \cite{Chialvo2012,Vescovi2016}.

\begin{figure*}
	\subfloat[$\phi_{\text{c}}=0.630\pm0.001$\label{fig:collapse_SigN-a}]{%
  		\includegraphics[width=0.5\textwidth]{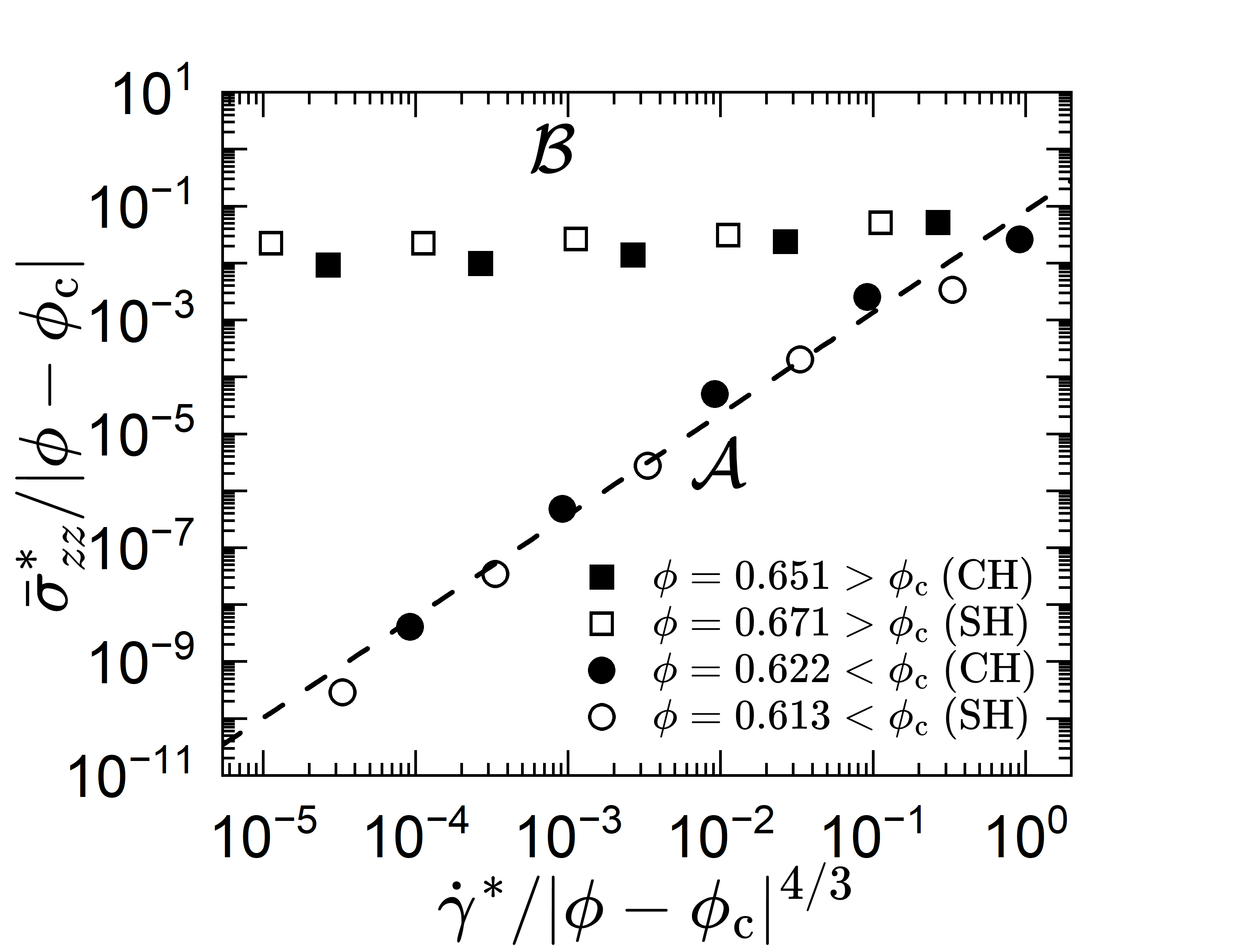}%
	}\hfill
	\subfloat[$\phi_{\text{c}}=0.585\pm0.002$\label{fig:collapse_SigN-b}]{%
  		\includegraphics[width=0.5\textwidth]{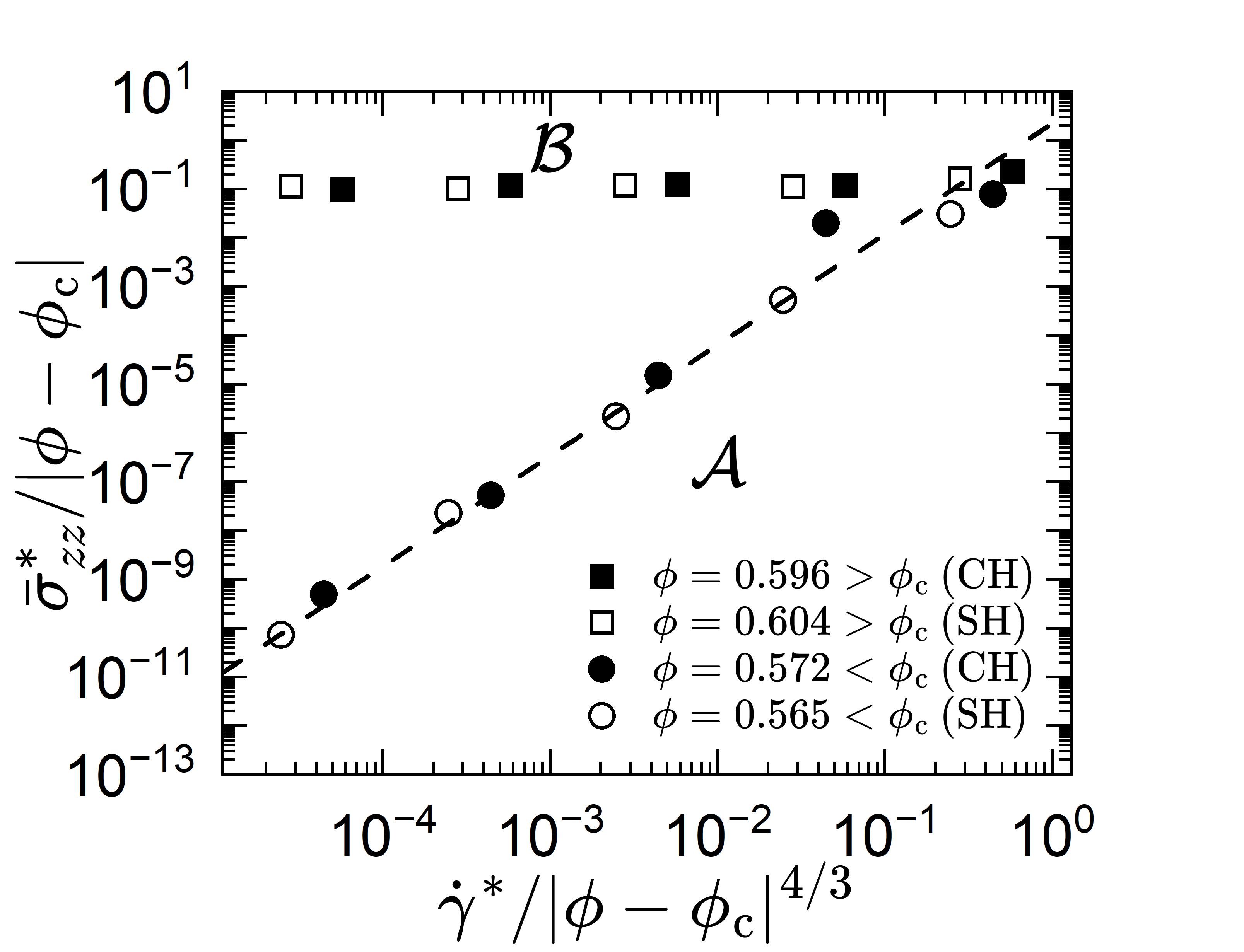}%
	}
	\caption{Time-averaged normal stresses $\bar{\sigma}_{zz}^*=\bar{\sigma}_{zz}/K_{\text{N}}$ versus shear rates $\dot{\gamma}^*=\dot{\gamma}d/\sqrt{K_{\text{N}}/\rho}$, both nondimensionalized, are acquired from the CH model (closed symbols) and the SH model (open symbols). (a) $\phi_{\text{c}}=0.630\pm0.001$. $\mu_0=0.001$ for the CH model and $V_{\text{c}}=6\times 10^{-5}$ cm/s (with $\lambda_{\text{SH}}=6\times 10^{-7}$) for the SH model. (b) $\phi_{\text{c}}=0.585\pm0.002$. $\mu_0=1.0$ for the CH model and $V_{\text{c}}=6000$ cm/s (with $\lambda_{\text{SH}}=60$) for the SH model. In each panel, the data are obtained with driving speeds from 0.01 cm/s up to 100 cm/s. After rescaled by $|\phi-\phi_{\text{c}}|^{\nu}$, the data in each panel are collapsed into two regimes: $\mathcal{A}$ ($\phi<\phi_{\text{c}}$, circular symbols) represents collisional flows in which a dashed line is fitted to the data from the CH model; $\mathcal{B}$ ($\phi>\phi_{\text{c}}$, square symbols) represents quasistatic flows in which $\bar{\sigma}_{zz}^*$ is almost independent of $\dot{\gamma}^*$. The slope of the dashed lines are $1.78\pm0.08$ and $2.27\pm0.15$ for the panels (a) and (b), respectively.}\label{fig:collapse_SigN}
\end{figure*}

In each panel of Fig.~\ref{fig:collapse_SigN}, we include two sets of simulations for both models, each of which contains data either above or below a similar value of $\phi_{\text{c}}$, i.e., $\phi_{\text{c}}=0.630\pm0.001$ and $0.585\pm0.002$, for Fig.~\ref{fig:collapse_SigN-a} and \ref{fig:collapse_SigN-b}, respectively. We plot the dimensionless time-averaged normal stress $\bar{\sigma}_{zz}^*=\bar{\sigma}_{zz}/K_{\text{N}}$ against the imposed dimensionless shear rate $\dot{\gamma}^*=\dot{\gamma}d/\sqrt{K_{\text{N}}/\rho}$, where $\rho$ is the mean mass density of the particles. Such dimensionless forms were pioneered by Campbell in his analyses on granular flows of soft particles using a linear contact model and various elasticity constants \cite{Campbell2002,Campbell2006}. Later these forms were adopted by Chialvo et al. \cite{Chialvo2012}, who rather focused on the role of the shear rate $\dot{\gamma}$ as we do in the present work. Following the scaling laws discovered by Chialvo et al. \cite{Chialvo2012}, we rescale both axes by factors $|\phi-\phi_{\text{c}}|^{\nu}$ such that $\nu=-1$ for $\bar{\sigma}_{zz}^*$ and $\nu=-4/3$ for $\dot{\gamma}^*$ in attempt to collapse the simulation results over various $\phi$ and material parameters. Two volume factions ($\phi$) are involved for data from either the CH or the SH model. These data are collapsed into two regimes: Regime $\mathcal{A}$ (with $\phi<\phi_{\text{c}}$) represents collisional flows, as the data fitting (dashed line) reveals scalings close to quadratic as $\bar{\sigma}_{zz}^*\propto\dot{\gamma}^{*2}$, implying the Bagnold scaling \cite{Bagnold1954,Lois2005,Forterre2008}. Regime $\mathcal{B}$ (with $\phi>\phi_{\text{c}}$) represents quasistatic flows, such that the time-averaged normal stress is almost independent of the dimensionless shear rate $\dot{\gamma}^*$.

\begin{figure*}
	\subfloat[$\phi_{\text{c}}=0.607\pm0.003$\label{fig:collapse_SigN-c}]{%
  		\includegraphics[width=0.5\textwidth]{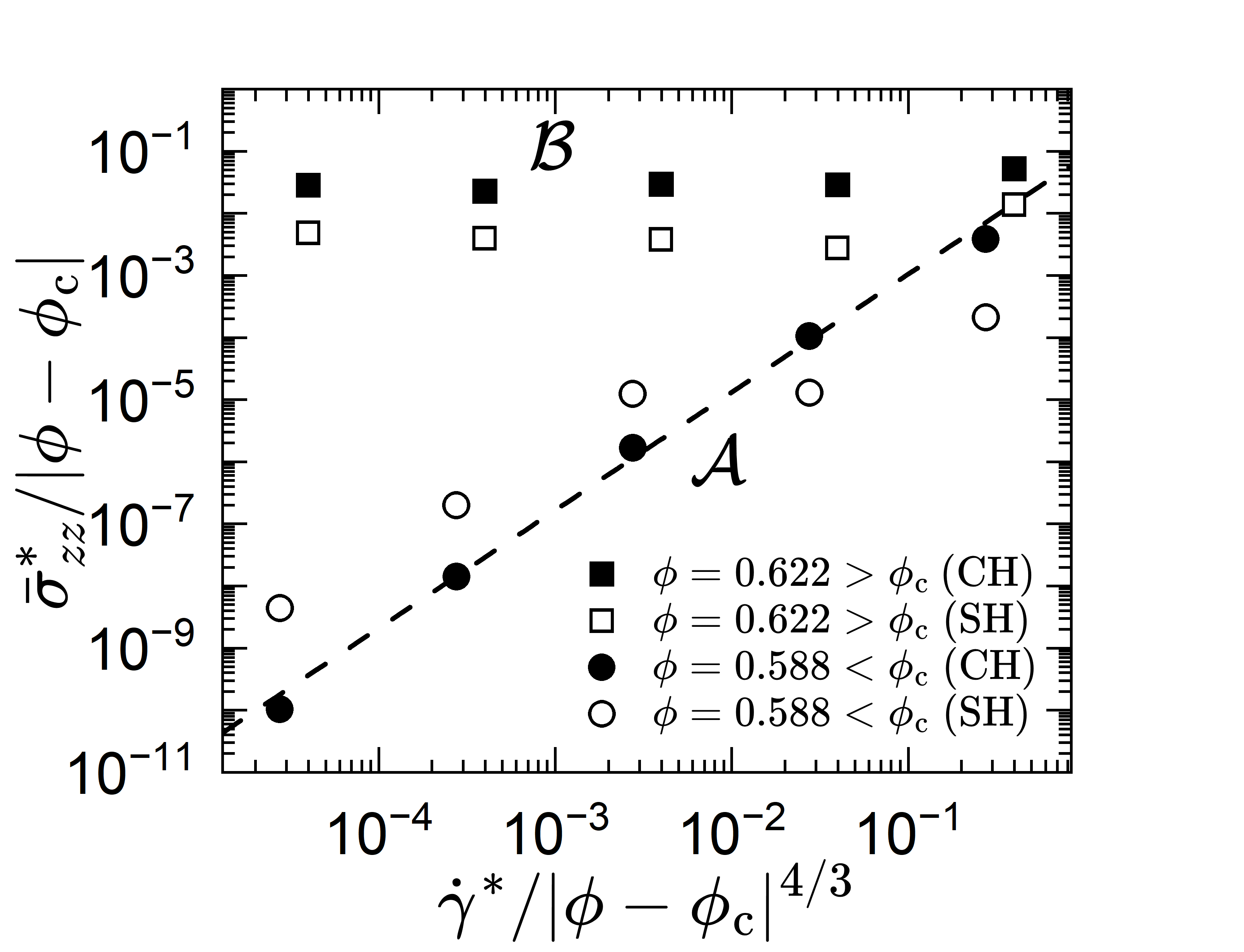}%
	}\hfill
	\subfloat[$\phi_{\text{c}}=0.586\pm0.003$\label{fig:collapse_SigN-d}]{%
  		\includegraphics[width=0.5\textwidth]{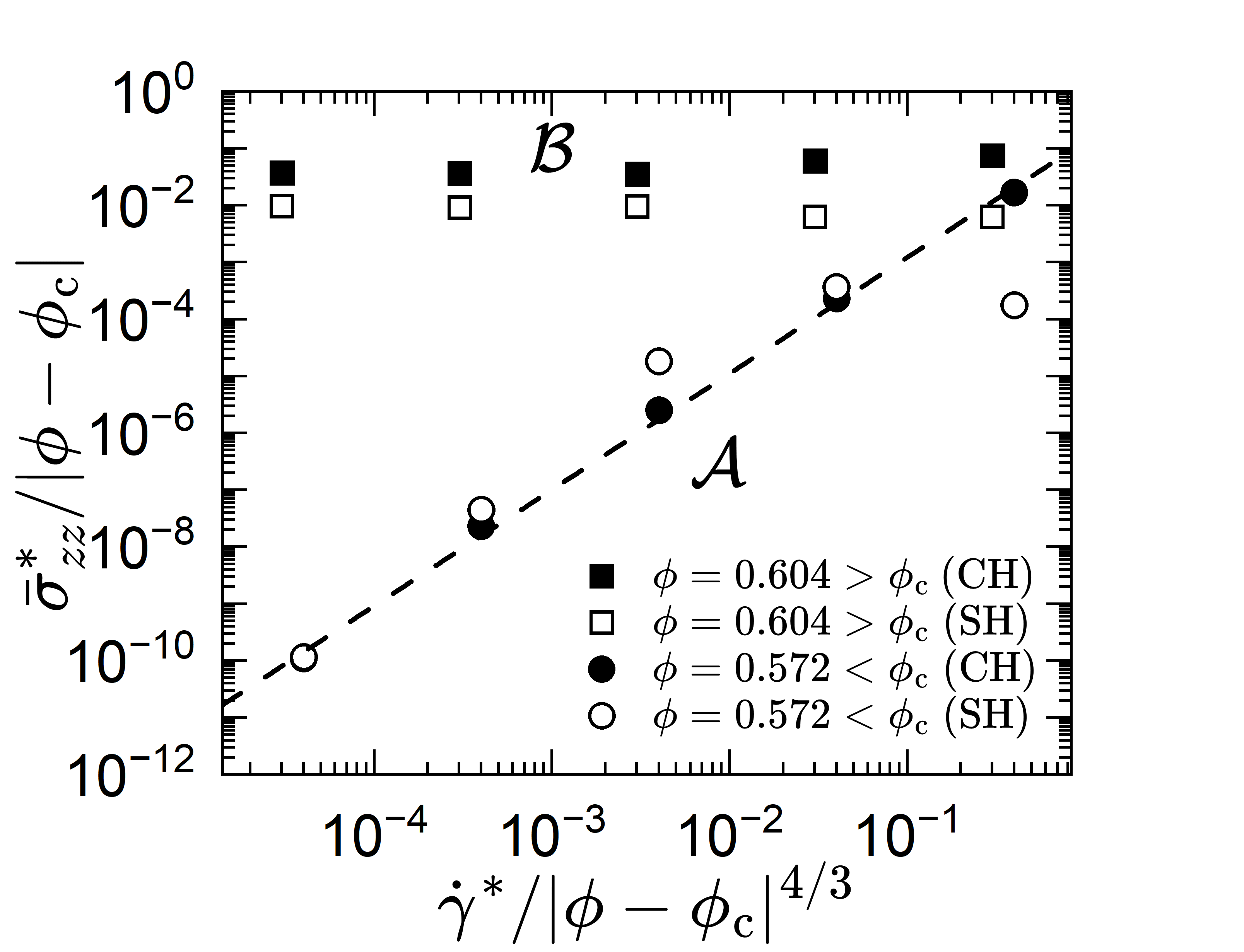}%
	}
	\caption{Same plot as Fig.~\ref{fig:collapse_SigN} except with the choices of $\lambda_{\text{SH}}$ that result in a clear departure from the data using the CH model, despite that $\phi_{\text{c}}$ being the same. (a) $\phi_{\text{c}}=0.607\pm0.003$. $\mu_0=0.2$ for the CH model and $V_{\text{c}}=0.06$ cm/s (with $\lambda_{\text{SH}}=6\times 10^{-4}$) for the SH model. (b) $\phi_{\text{c}}=0.586\pm0.003$. $\mu_0=0.5$ for the CH model and $V_{\text{c}}=0.6$ cm/s (with $\lambda_{\text{SH}}=6\times 10^{-3}$) for the SH model. The slope of the dashed lines are $1.90\pm0.07$ and $2.04\pm0.05$ for the panels (a) and (b), respectively.}\label{fig:collapse_SigN_2}
\end{figure*}

Figures~\ref{fig:collapse_SigN-a} and \ref{fig:collapse_SigN-b} give two examples that, in terms of the time-averaged rheology, a granular packing governed by the SH model can be indistinguishable from that governed by the CH model. Despite the different values of $\phi$ for the CH and the SH model in each panel, these data points are all collapsed into the two regimes after the rescaling by the distance to the shared value of $\phi_{\text{c}}$. This has confirmed that mapping the dynamics governed by the CH and SH model based on the distance to $\phi_{\text{c}}$ is indeed possible. On the other hand, Figs.~\ref{fig:collapse_SigN-c} and \ref{fig:collapse_SigN-d} demonstrate that, with some certain values of $\lambda_{\text{SH}}$ (varied by changing $V_{\text{c}}$), data of the SH model exhibit a clear deviation from those of the CH model, even with the similar values of $\phi_{\text{c}}$ and the same distance to these $\phi_{\text{c}}$.


\subsection{Time-averaged rheology and the temporal fluctuation}
In order to gain insight into how the deviation from that of the CH model affect the macroscopic properties of the SH model, we further compare the rheological responses from both models. For a steady flow, it is common to characterize the rheology by expressing the stress ratio $\mu_{\text{eff}}\equiv \bar{\sigma}_{xz}/\bar{\sigma}_{zz}$ as a function of the inertial number $I\equiv\dot{\gamma}d/\sqrt{\bar{\sigma}_{zz}/\rho}$. Here, $\bar{\sigma}_{xz}$ (or $\bar{\sigma}_{zz}$) is the shear (or normal) stress on the wall and the bar stands for the time average. This ratio is often related to the inclination of the force chains relative to the boundaries \cite{Cates1998,Aharonov2004,Barreto2012}. In the conventional view (the CH model), at the low inertial number limit, the ratio is found to stay at a finite value that depends on the Coulomb coefficient $\mu_0$ \cite{Midi2004,Berzi2024}. Such $\mu_{\text{eff}}$-$I$ relationship is proved to be generic for both rigid \cite{Midi2004} and soft particles \cite{Chialvo2012,Gu2016} under simple shear and other configurations \cite{Jop2006}.

\begin{figure}
	\includegraphics[width=\columnwidth]{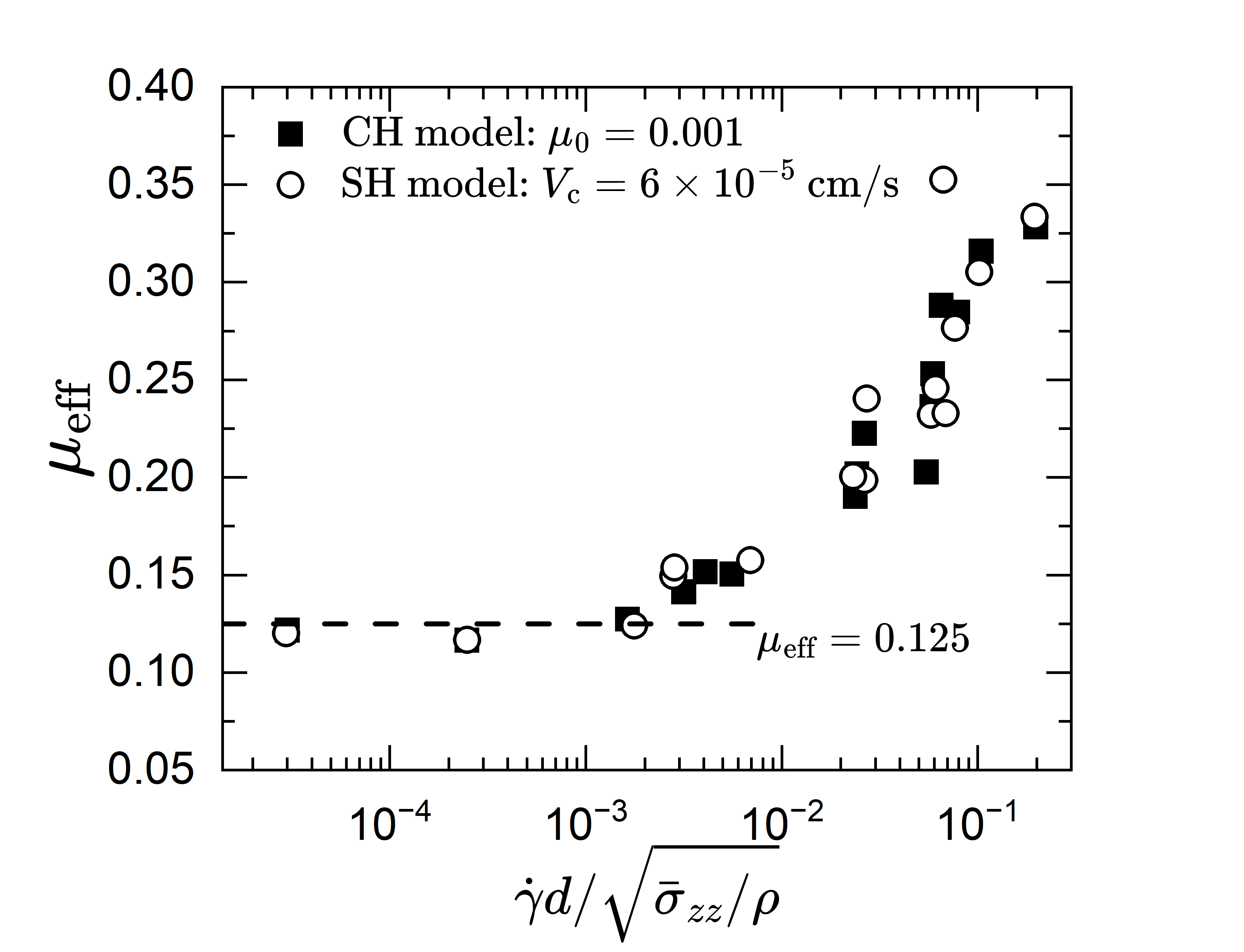}
	\caption{The stress ratio $\mu_{\text{eff}}$ plotted against the inertial number $I=\dot{\gamma}d/\sqrt{\bar{\sigma}_{zz}/\rho}$, for granular packings governed by the CH and the SH model, respectively, that share similar values of $\phi_{\text{c}}$. $\phi_{\text{c}}=0.630\pm0.001$, with $5.711\times 10^{-4} \:\text{s}^{-1}\le\dot{\gamma}\le5.972\:\text{s}^{-1}$ and $0.613\le\phi\le 0.641$. The dashed line indicates the stable stress ratio in the low inertial number limit. Both models exhibit identical rheological behavior over a wide range of $I$, as expected based on the universal behavior shown in Fig.~\ref{fig:collapse_SigN-a}.}\label{fig:mueff_I-a}
\end{figure}

Figure~\ref{fig:mueff_I-a} demonstrates the case with $\phi_{\text{c}}=0.630\pm0.001$ ($V_{\text{c}}=6\times 10^{-5}$ cm/s for the SH model and $\mu_0=0.001$ for the CH model) that is associated with those data in Fig.~\ref{fig:collapse_SigN-a}. The graph shows a large collection of data with different combinations of $\dot{\gamma}$ and $\phi$. Data from both models collapse on the same curve. In other words, the SH model and the CH model are indistinguishable in terms of the time-averaged rheology. And results from both models indicate a stable stress ratio (0.125) in the low inertial number limit. 

\begin{figure*}
	\centering
	\begin{tabular}{cc}
	\adjustbox{valign=b}{\subfloat[\label{fig:mueff_I-b}]{%
    	\includegraphics[width=0.7\textwidth]{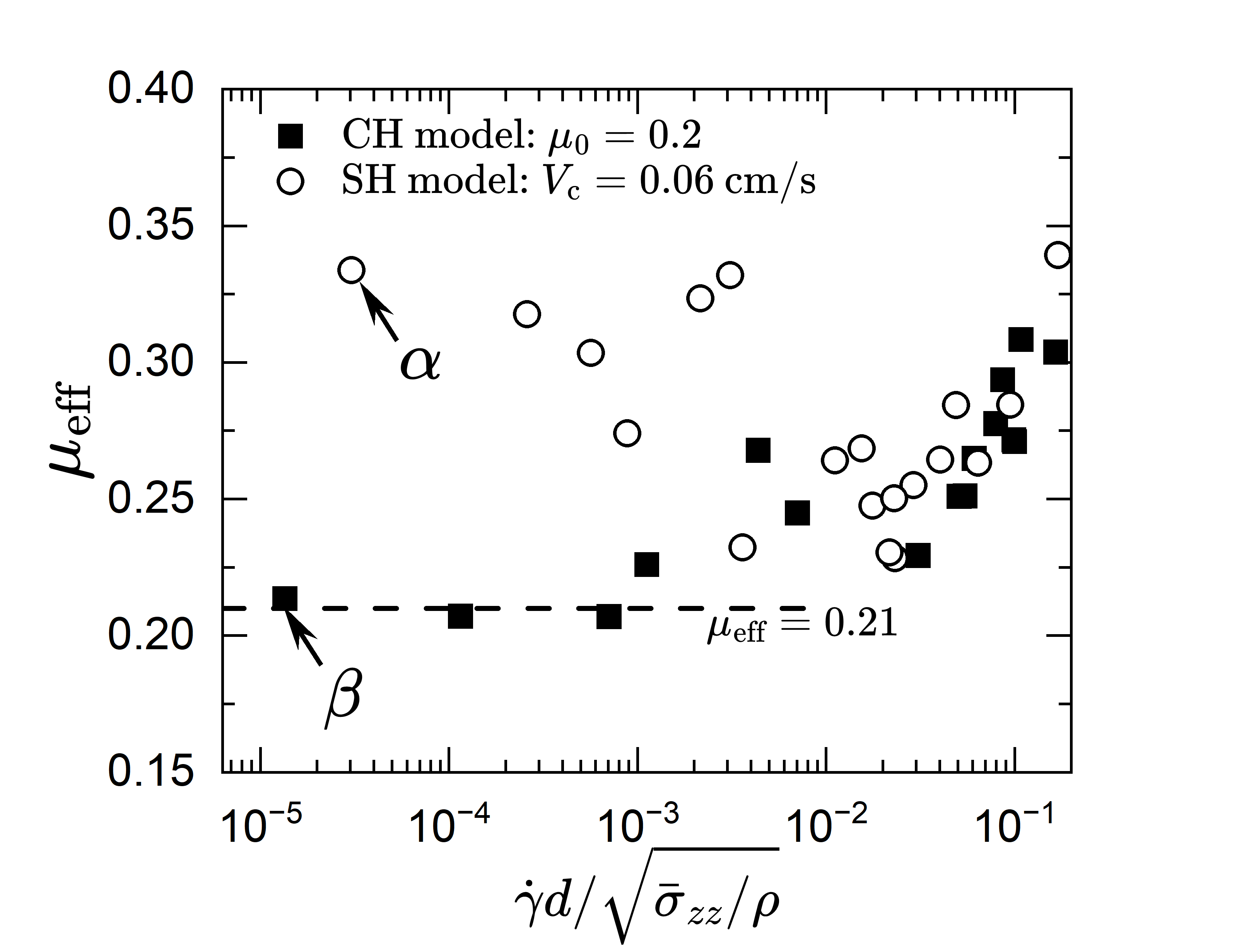}}}
          &
	\adjustbox{valign=b}{\begin{tabular}{@{}c@{}}
    \subfloat[$\alpha$, $I=3.05\times 10^{-5}$\label{fig:PS_timeseries-a}]{%
          \includegraphics[width=0.3\textwidth,height=0.225\textwidth]{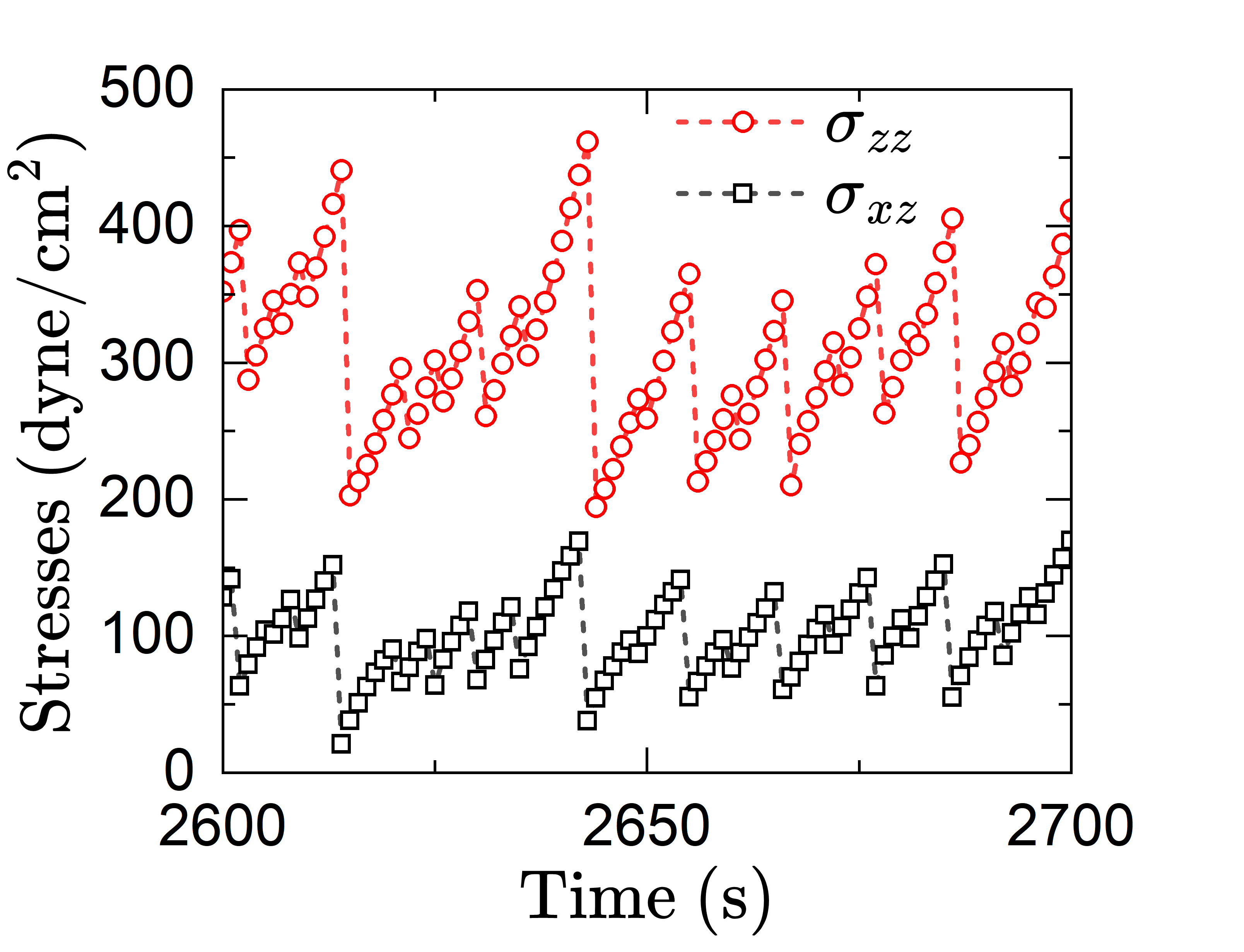}} \\
    \subfloat[$\beta$, $I=1.35\times 10^{-5}$ \label{fig:PS_timeseries-b}]{%
          \includegraphics[width=0.3\textwidth,height=0.225\textwidth]{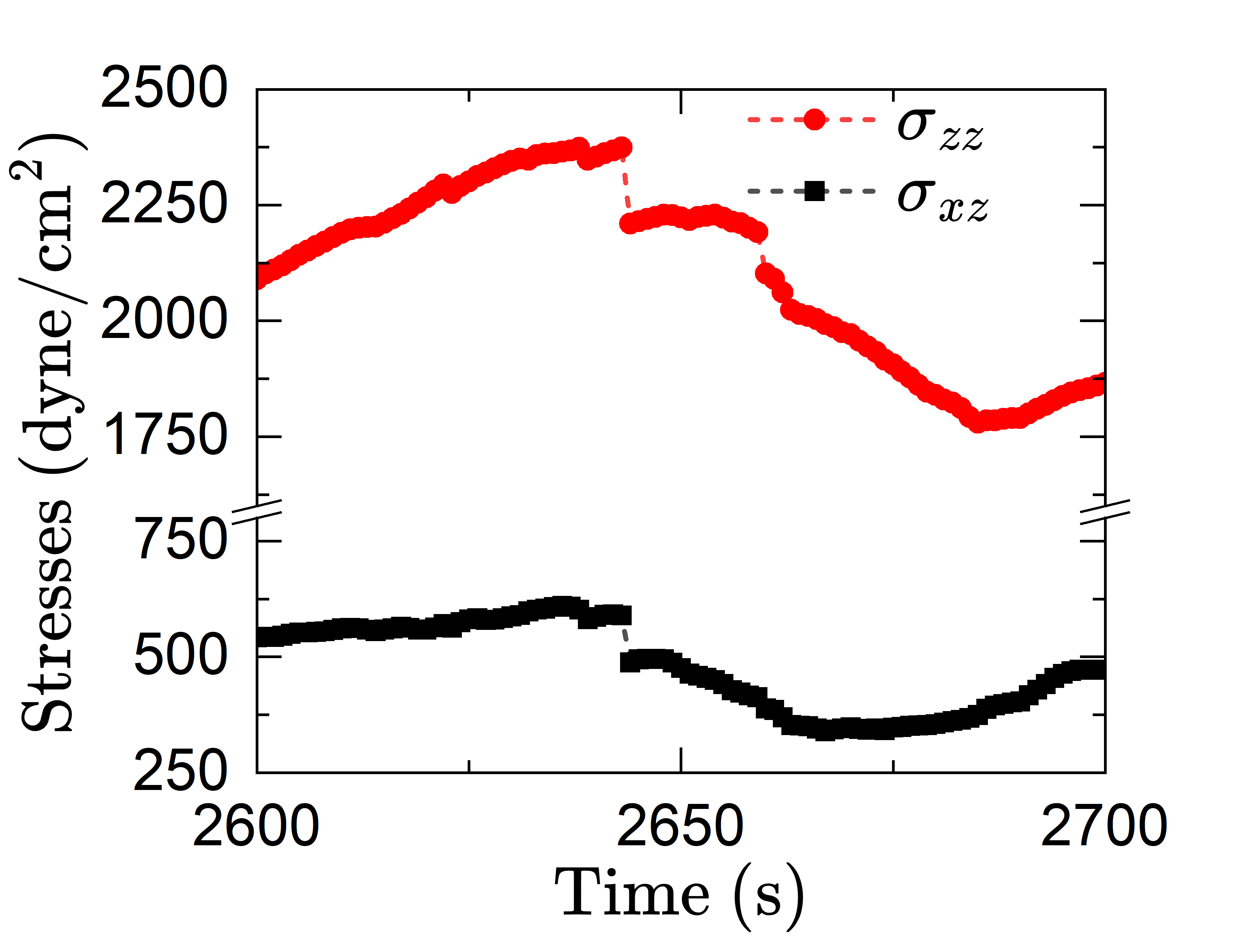}}
	\end{tabular}}
	\end{tabular}
	\caption{(a) Same plot as Fig.~\ref{fig:mueff_I-a} except that $\phi_{\text{c}}=0.607\pm 0.003$ with $4.994\times 10^{-4} \:\text{s}^{-1}\le\dot{\gamma}\le 5.711\:\text{s}^{-1}$ and $0.536\le\phi\le0.613$. While both models follow the same curve for $I>2\times 10^{-2}$, data from the SH model deviate substantially from the stable stress ratio (dashed line) at the low inertial number limit. Panels (b) and (c) are time sequences of the instantaneous stresses $\sigma_{xz}$ and $\sigma_{zz}$ extracted from the simulation data corresponding to points $\alpha$ and $\beta$ in the panel (a), respectively. The deviation of the open circles in Panel (a) corresponds to the occurrence of quaking as shown in Panel (b).}
\end{figure*}

For comparison, Fig.~\ref{fig:mueff_I-b} demonstrates the case with $\phi_{\text{c}}=0.607\pm0.003$ ($V_{\text{c}}=0.06$ cm/s for the SH model and $\mu_0=0.2$ for the CH model) that is associated with those data in Fig.~\ref{fig:collapse_SigN-c}. It is worth noting that, for inertial numbers $I<2\times 10^{-2}$, the data from the SH model no longer overlap with those from the CH model. In other words, in the quasistatic regime, the dynamics governed by the SH model substantially changes from that of the CH model.

In Figs.~\ref{fig:PS_timeseries-a} and \ref{fig:PS_timeseries-b}, we plot the values of instantaneous stresses as functions of time extracted from the simulation data corresponding to the point $\alpha$ (governed by the SH model) and the point $\beta$ (governed by the CH model) indicated by the arrows in Fig.~\ref{fig:mueff_I-b}; both are in the quasistatic regime. Figure~\ref{fig:PS_timeseries-a} exhibits quasi-periodic growths and sudden drops of stresses, to which we refer as quaking \cite{Tsai2024b}. In contrast, Fig.~\ref{fig:PS_timeseries-b} shows relatively smooth curves and occasional small drops. 

Comparing Figs.~\ref{fig:PS_timeseries-a} and \ref{fig:PS_timeseries-b}, we find that the typical magnitude of stresses in Fig.~\ref{fig:PS_timeseries-b} is almost five times that of Fig.~\ref{fig:PS_timeseries-a}, even if their volume fractions are exactly the same. This suggests that the repeated reorganizations of particles (the quaking events in the SH simulations) have significantly weakened the granular packing. 

\subsection{State diagrams at different driving speeds}
\begin{figure*}
	\subfloat[$U=0.01$ cm/s\label{fig:phasediag_heat-a}]{%
  		\includegraphics[width=0.306\textwidth]{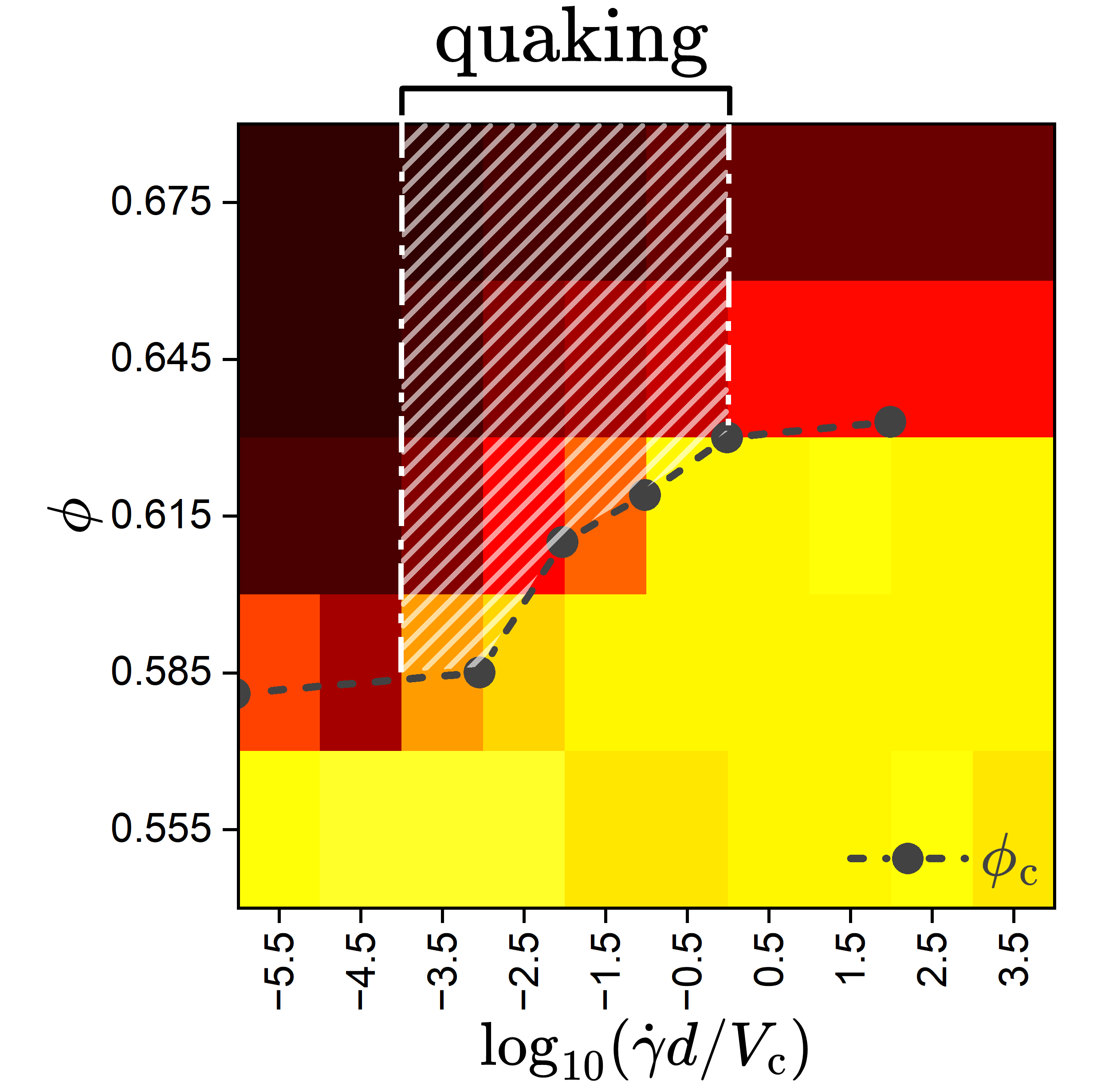}%
	}\hfill
	\subfloat[$U=1.0$ cm/s\label{fig:phasediag_heat-b}]{%
  	\includegraphics[width=0.306\textwidth]{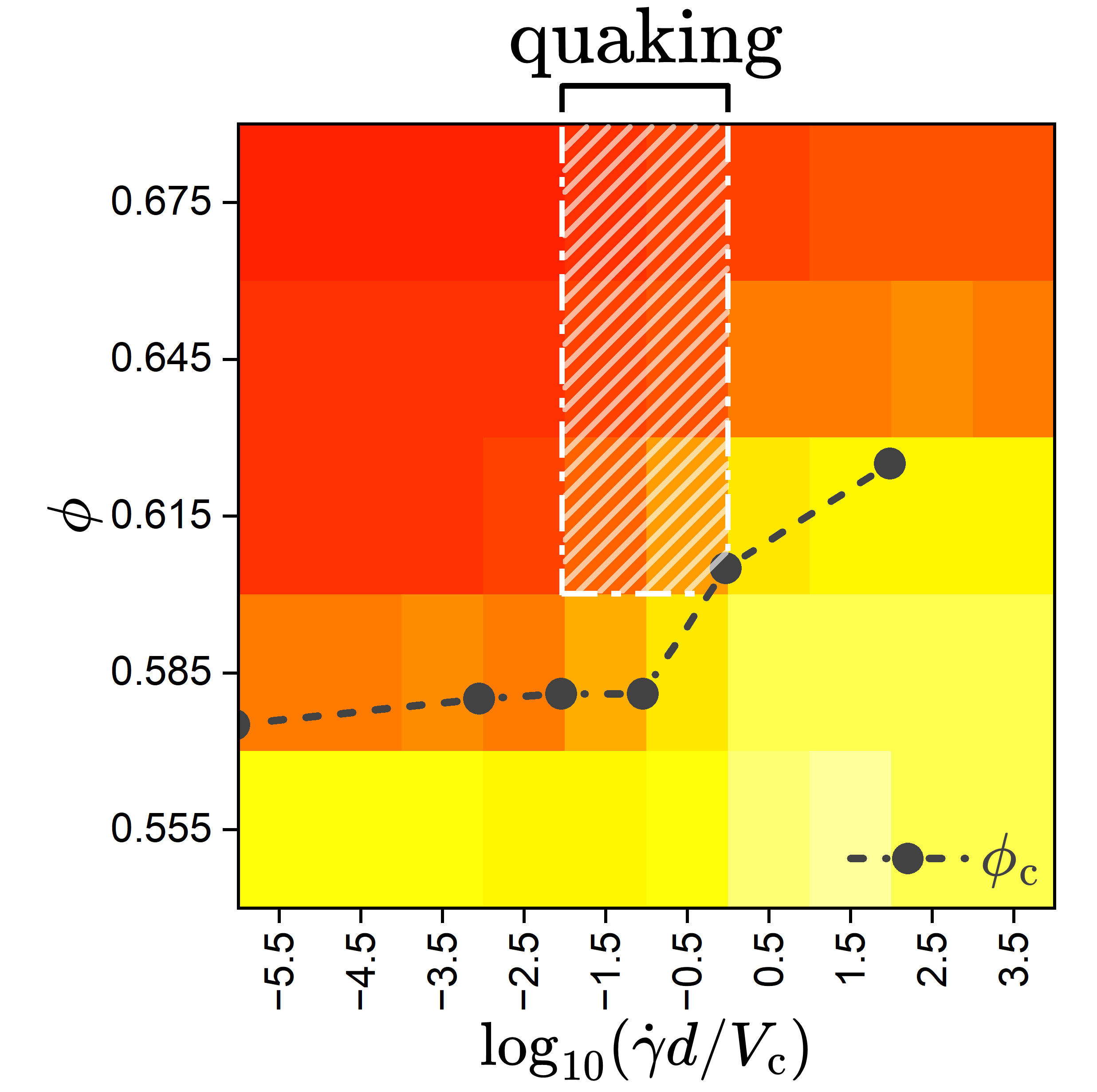}%
	}\hfill
	\subfloat[$U=100$ cm/s\label{fig:phasediag_heat-c}]{%
  		\includegraphics[width=0.388\textwidth]{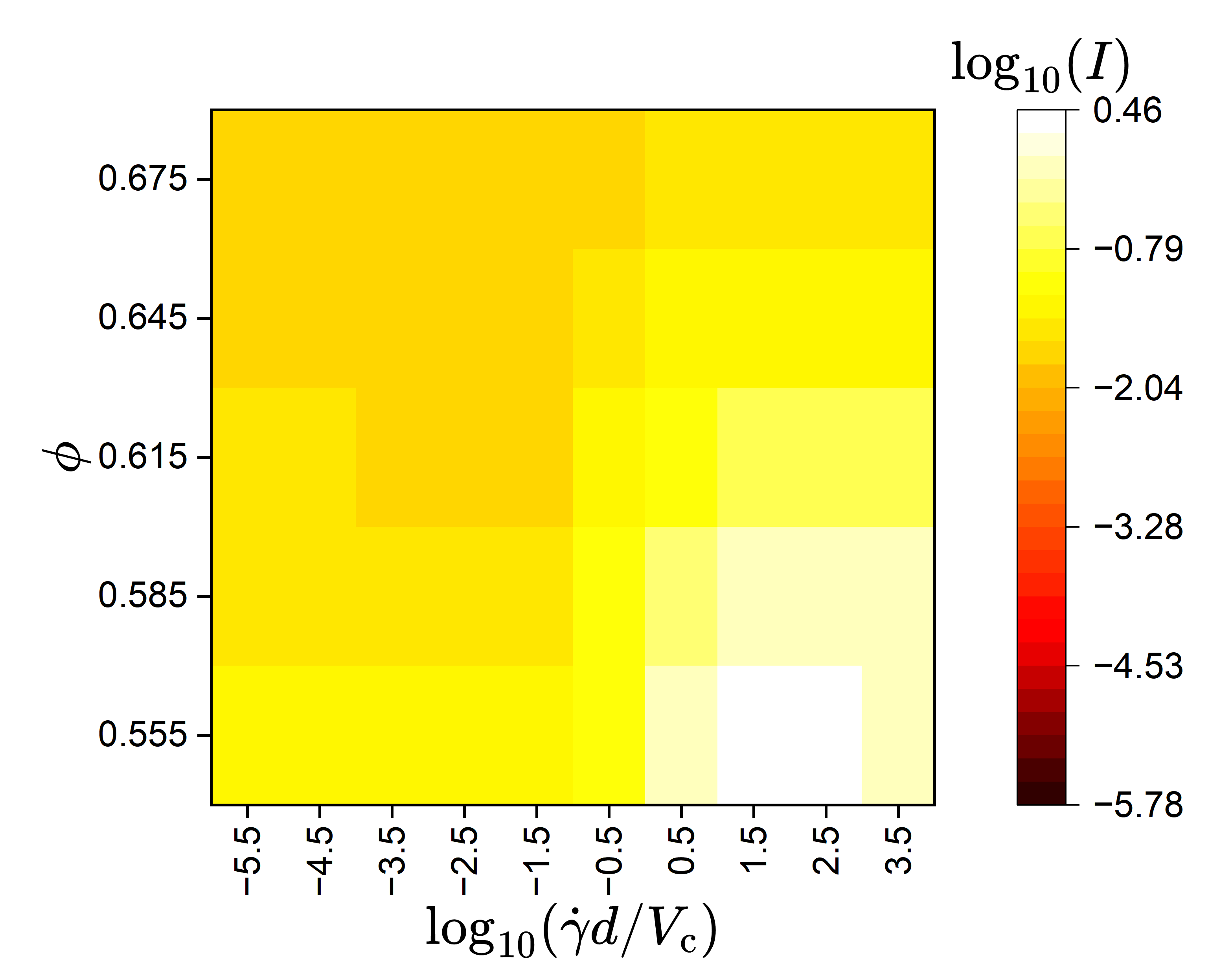}%
	}
	\caption{State diagrams for granular packings governed by the SH model, at three different driving speeds $U$. The data include simulations from thirteen different volume fractions $\phi$ but presented as five bins along the vertical direction. The horizontal coordinate is the logarithm of the slipperiness $\dot{\gamma}d/V_{\text{c}}$, generated from ten different values of $V_{\text{c}}$. The color shows the logarithm of the inertial number $I=\dot{\gamma}d/\sqrt{\bar{\sigma}_{zz}/\rho}$. In the panels (a) and (b), the closed circles indicate the locations of the critical volume fraction $\phi_{\text{c}}$.}\label{fig:phasediag_heat}
\end{figure*}

In Fig.~\ref{fig:phasediag_heat}, we present the state diagrams for granular packings governed by the SH model at three different driving speeds, $U=0.01$ cm/s, 1.0 cm/s and 100 cm/s, respectively. (In our previous work \cite{Tsai2024b}, the granular packing was driven at $U=0.1$ cm/s, with all other conditions identical to the present work.) The vertical and horizontal axes are respectively the volume fraction $\phi$ and the logarithm of the slipperiness $\dot{\gamma}d/V_{\text{c}}$ \cite{Tsai2024b}. In this two-dimensional parameter space, the value of inertial number $I=\dot{\gamma}d/\sqrt{\bar{\sigma}_{zz}/\rho}$ is computed and color-coded like a heat map, based on various values of $\phi$ and $V_{\text{c}}$ (across ten decades). The quaking regime is identified (and indicated by stripes) by observing the time sequence of the instantaneous coordination number, the elastic energy density, or the boundary stresses, since the occurrence of stick-slip fluctuations are consistent in time.

A very notable feature is that, as the value of $U$ increases, the quaking regime suffers a shrinkage and eventually disappears from the parameter space. This can be perceived as the consequence of the increase of the inertial numbers: The entire parameter space eventually becomes collisional everywhere, so that the speed-dependent friction in the SH model can no longer produce the instability for triggering the quaking events. This is also consistent with Figs.~\ref{fig:collapse_SigN} and \ref{fig:collapse_SigN_2}, which show that the two \emph{branches} (quasistatic versus collisional) are progressively merging as the shear rate increases. In Figs.~\ref{fig:phasediag_heat-a} and \ref{fig:phasediag_heat-b}, we have also indicated the location of $\phi_{\text{c}}$. Note that the curve of $\phi_{\text{c}}$ also serves as the lower boundary of the quaking regime.

\section{Discussion}\label{sec:discussion}
In Sec.~\ref{sec:phi_c}, we have interpreted $\lambda_{\text{SH}}$ as the ratio of the retardation time $\tau_{\text{R}}=m\gamma_{\text{N}}/K_{\text{N}}$ to the critical weakening time $\tau_{\text{c}}=d/V_{\text{c}}$. Since these particles are characterized by the viscoelastic contact model Eq.~(\ref{eq:normal_force}), the elastic response in the normal direction between two colliding particles is not instantaneous but delayed, of which the timescale depends on $\gamma_{\text{N}}$. Specifically, our $\gamma_{\text{N}}$ is sufficiently large such that the system is overdamped. Note that the terms in the parentheses of Eq.~(\ref{eq:normal_force}) is analogous to the Kelvin-Voigt material, the simplest model for viscoelastic solids that is composed of a linear spring (characterized by the elastic modulus $E$) and a viscous damper (characterized by the viscosity $\eta$) linked in parallel \cite{Hajikarimi2021}; this model is overdamped by nature as no inertial term is present. The Kelvin-Voigt material has only one characteristic timescale, $\tau_{\text{KV}}=\eta/E$.

In our SH model, the retardation time $\tau_{\text{R}}=m\gamma_{\text{N}}/K_{\text{N}}$ is analogous to $\tau_{\text{KV}}$ in the Kelvin-Voigt material. Essentially, $\tau_{\text{KV}}$ signifies how much time the Kelvin-Voigt material will take to reach the strain responding to a constant applied stress. Therefore, unless there is a network of enduring contacts ($\phi>\phi_{\text{c}}$), $\tau_{\text{R}}$ is a crude, simple estimate of the duration in which \emph{two particles} transiently overlap while they are sliding with each other. On the other hand, when the sliding speed between two particles exceeding $V_{\text{c}}$, the critical weakening time $\tau_{\text{c}}$ is the longest time in which two overlapped particles sliding through a relative displacement equivalent to the mean diameter $d$. Therefore, one might expect that the velocity weakening should occur when $\lambda_{\text{SH}}\lesssim 1$. However, our results in Fig.~\ref{fig:phic-Vc-b} indicate a lower threshold down to the order of $10^{-2}$. The reason behind such discrepancy remains elusive and requires further investigation in the future.

Also note that the resemblance of Figs.~\ref{fig:phic-Vc-a} and \ref{fig:phic-Vc-b} by no means suggests that $\lambda_{\text{SH}}$ is an apparent friction coefficient. By drawing the analogy between $\tau_{\text{R}}$ and $\tau_{\text{KV}}$, we can see why $\lambda_{\text{SH}}$ is relevant only close to $\phi_{\text{c}}$ by two conditions. First, the system must be overdamped. Particle collisions are underdamped when $\phi$ is far below $\phi_{\text{c}}$; in these scenarios the relevant timescale shall be the collision time \cite{Brilliantov1996} or simply $\dot{\gamma}^{-1}$.

Second, the overlapping is transient such that $\tau_{\text{R}}$ is relevant. The fluctuation of coordination numbers is at the peak when close to $\phi_{\text{c}}$ as shown in Fig.~\ref{fig:find_phic}, suggesting frequent changes in overlapping between the neighboring particles. Once the force network is formed above $\phi_{\text{c}}$, however, $\tau_{\text{R}}$ no longer represents the timescale of sliding due to the multiple, enduring overlaps per particle (i.e., sticking is dominant over sliding), and thus $\lambda_{\text{SH}}$ is also irrelevant in such scenarios.

Finally, the two conditions suggest that $\phi_{\text{c}}$ of the SH model can substantially shift according to the choice of the contact model and the material parameters such as $K_{\text{N}}$, $\gamma_{\text{N}}$, and so on.

\section{Summary and Concluding Remarks}
In the present work with numerical simulations, we drive the granular packing beyond the quasistatic regime, and make detailed comparisons of results from the previously proposed Stribeck-Hertz model (SH, with a speed-dependent friction) and from the conventional Coulomb-Hertz model (CH, with a constant friction coefficient).
The simulations from both models show that the values of the critical volume fraction $\phi_{\text{c}}$ are determined by the material parameters only, independent of the driving speed. Under both models in certain combinations of material parameters, the relationships between the mean normal stress and the shear rate can collapse if the differences between the volume fraction ($\phi$) and the critical volume fraction ($\phi_{\text{c}}$) are properly scaled. 
Meanwhile, we find that quaking is a distinct feature of the SH model: This is reflected by the anomalous stress ratio at low inertial numbers and the time sequences of instantaneous stresses. Investigations into the stress responses at various driving speeds make us conclude that the quaking occurs only in the intermediate range of slipperiness, in consistence with our previous work, but further reveal the progressive narrowing of the quaking regime and the disappearance of it at an extremely high shear rate.

From this work, we find that the volume fraction going above the critical volume fraction $\phi_{\text{c}}$ is a necessary condition for the quaking to occur. The effect of grain inertia generally suppresses the occurrence of quaking. And we expect that the ability to apply the SH model with the analysis of the critical volume fraction to a laboratory setting or field observations such as debris flows \cite{Kostynick2022} will help us better understand the nature of granular flows in frictional grains.





\bibliographystyle{elsarticle-num} 
\bibliography{ref-elsarticle-ver32}




\end{document}